\documentclass[prb, aps, 10pt, twocolumn, superscriptaddress, nofootinbib]{revtex4-2}
\usepackage[ascii]{inputenc}
\usepackage{amsmath,amssymb,amsfonts,amsthm}
\usepackage{mathtools}
\usepackage{braket}
\usepackage{tikz}
\usepackage{siunitx}
\usetikzlibrary{calc, decorations.pathreplacing, decorations.pathmorphing, 3d, patterns}
\usepackage{orcidlink}
\usepackage{graphicx}
\usepackage{hyperref}
\usepackage{listings}               
\definecolor{dkgreen}{rgb}{0,0.6,0}
\definecolor{gray}{rgb}{0.5,0.5,0.5}
\definecolor{mauve}{rgb}{0.58,0,0.82}
\definecolor{back}{rgb}{0.96,0.96,0.96}

\DeclarePairedDelimiter\norm{\lVert}{\rVert}
\DeclarePairedDelimiter\abs{\lvert}{\rvert}

\DeclareMathOperator*{\argmin}{argmin}
\DeclareMathOperator{\tr}{tr}
\newcommand{\cntr}{\mathcal{C}}

\tikzset{
    varnode/.style = {circle, draw=black, minimum size=10pt}, 
    facnode/.style = {rectangle, draw=black, minimum size=10pt}, 
    kronnode/.style = {circle, draw=black, fill=black, inner sep=0pt, minimum size=5pt}, 
    hadamardnode/.style = {rectangle, fill=yellow, draw=black, minimum size=10pt},
    tpsbase/.style = {draw=black, fill=blue!70, inner sep=0pt, minimum size=5pt}, 
    tpsnode/.style = {circle, tpsbase}, 
    tpobase/.style = {draw=black, fill=yellow!70, inner sep=0pt, minimum size=5pt}, 
    tponode/.style = {circle, tpobase},
    msgbase/.style = {draw=black, fill=blue!40, inner sep=0pt, minimum size=5pt}, 
    msgnode/.style = {rectangle, msgbase}, 
    projnode/.style = {rectangle, draw=black, fill=green!60, inner sep=0pt, minimum size=5pt}, 
    blankmsgnode/.style = {},
    physedge/.style = {green!75!black}, 
    excedge/.style = {red, very thick}, 
    highlightbox/.style = {fill=magenta, opacity=.1, dashed}, 
    >=stealth,
}

\begin{document}

\title{DMRG using Belief Propagation}

\author{Hendrik K{\"u}hne\orcidlink{0009-0000-7097-3762}}
\email{hendrik.kuehne@tum.de}
\affiliation{Technical University of Munich, School of Natural Sciences, Department of Physics, James-Franck-Stra{\ss}e 1, 85748 Garching, Germany}

\author{Christian B.~Mendl\orcidlink{0000-0002-6386-0230}}
\email{christian.mendl@tum.de}
\affiliation{Technical University of Munich, School of Computation, Information and Technology, Department of Computer Science, Boltzmannstra{\ss}e 3, 85748 Garching, Germany}
\affiliation{Technical University of Munich, Institute for Advanced Study, Lichtenbergstra{\ss}e 2a, 85748 Garching, Germany}

\date{July 2026}

\begin{abstract}
    Tensor networks have attracted much attention as a powerful tool for modeling quantum many-body systems. Their contraction is a significant challenge, however, especially in highly connected networks, as memory requirements become prohibitive and the optimal contraction order is increasingly hard to find. The belief propagation (BP) algorithm has emerged as an alternative to exact contraction. Being formulated in a graph-agnostic way, it offers great flexibility, but its accuracy suffers in the presence of loops. In this work, we combine BP with the DMRG algorithm to solve ground-state problems, thereby extending DMRG to higher dimensions and arbitrary lattices. We demonstrate the viability of BP-DMRG on the transverse-field Ising model on a $2\times 2$ hexagonal lattice, finding that it produces states with a fidelity between $0.9$ and $0.99$ to the true ground state, and energy estimates with a relative error between $10^{-2}$ and $10^{-3}$. Additionally, BP-DMRG can find ground states on randomly generated lattices, with fidelity improving as the transverse field increases. We conclude with a discussion of the limitations we encounter when using belief propagation, highlighting that the TFI tensor network operators lead to larger errors during BP iterations in BP-DMRG.
\end{abstract}

\maketitle

\section{Introduction}
\label{sec:intro}

In recent years, computational physics has witnessed many advances driven by tensor networks (TNs)~\cite{Bridgeman:2017:InterpretativeDance, Berezutskii:2025:TN_for_quantum, Ferris:2014:TNS_QEC, Garcia:2024:TensorNetworkApplications, Cirac:2021:MPS_PEPS, Banuls:2023:TensorNetworkAlgorithms, FelserMontangero:2026:TensorNetworks}, since TNs offer ways to circumvent or, at least, mitigate the ``curse of dimensionality''. Many powerful methods for the simulation of quantum many-body physics (QMBP) and quantum computing have recently been proposed~\cite{Begusic:2024:FastConvergedClassicSim, Chen:2023:SmallEntanglementQFT, Schollwoeck:2011:DMRG, Stoudenmire:2012:TwoDimDMRG, Tindall:2024:EfficientEagleSim}, demonstrating that they can match the computational power that was thought to be granted only by quantum computers, e.g., in the context of quantum supremacy claims or simulations of the kicked Ising model~\cite{Arute:2019:QuantumSupremacy, Gray:2021:HyperContraction, Kim:2023:QuantumUtility, Begusic:2024:FastConvergedClassicSim, Tindall:2024:EfficientEagleSim}.

The belief propagation (BP) algorithm has emerged as an important technique in quantum computing and quantum information in this context~\cite{Alkabetz:2021:TN_BP, Begusic:2024:FastConvergedClassicSim, Mueller:2025:BP_memory_decoding, Tindall:2023:BP_gauging, Tindall:2024:EfficientEagleSim}. Originating from statistical inference on graphical models~\cite{Koller:2009:ProbabilisticGraphicalModels, Mezard:2009:InformationPhysicsComputation}, BP has garnered much attention as a tool for computationally efficient contraction of large TNs~\cite{Begusic:2024:FastConvergedClassicSim, Mezard:2009:InformationPhysicsComputation}. Contracting a TN is a key step in TN-based algorithms~\cite{Berezutskii:2025:TN_for_quantum, Bridgeman:2017:InterpretativeDance, Ferris:2014:TNS_QEC, Orus:2014:MPS_PEPS:intro}, but it poses significant computational challenges~\cite{Geiger:2025:OptimizingPartitioning, Gray:2024:HyperApproxContraction, Gray:2021:HyperContraction, Pancotti:2023:1RSP, Pfeifer:2014:OptContractionSequences}. The advantage of BP is its memory efficiency and applicability to arbitrary TN topologies; nevertheless, it represents an uncontrolled approximation in the presence of loops in the TN graph~\cite{Alkabetz:2021:TN_BP, Evenbly:2026:LoopSeriesExpansionPublished}.

\vspace{\baselineskip}

This work investigates the utility of BP for the ground-state problem. The gold standard in this field is the density matrix renormalization group algorithm (DMRG), a variational method for the optimization of tensor network states (TNS)~\cite{Schollwoeck:2011:DMRG, Verstraete:2023:30yearsDMRG, Dukelsky:1998:DMRGinMPS}. Applying DMRG on higher-dimensional or less-structured lattices remains challenging due to a lack of canonical forms, which incurs algorithmic and numerical challenges~\cite{Stoudenmire:2012:TwoDimDMRG}.

We propose a new variant of DMRG, called BP-DMRG, which uses BP within DMRG to extend the latter's applicability to arbitrary lattices. We benchmark it on the transverse-field Ising model (TFI), focusing on the extent to which the ground state problem (and BP itself) can be tackled in a lattice-agnostic way, thereby highlighting the opportunities and shortcomings of BP in the context of QMBP.

%

\section{Background}
\label{sec:backgr}

Before introducing the BP-DMRG algorithm, we will devote the following two sections to its two main building blocks: belief propagation and DMRG.

\subsection{Belief Propagation}
\label{sec:backgr:BP}

We shall first introduce BP for TN contraction, and then turn to tensor network states and operators.

\subsubsection{Tensor Network Contraction}
\label{sec:backgr:BP:TN}

Consider the problem of TN contraction. Let $(V, E)$ be a graph, where to each node $a\in V$ one associates a tensor $T_a$. There are $|E|$ edges in total; if $(a,b)\in E$, the tensors $T_a$ and $T_b$ share the index $i_{(a,b)}$. TN contraction is then the task of summation over all shared indices, until a scalar $Z$ remains:
\begin{equation}
    \begin{gathered}
        Z = \sum_{\{i_{(a,b)}\}} \prod_{c\in V}T_c^{[\xi_c]} \equiv \cntr\left\{\prod_{a\in V}T_a\right\}, \\
        \text{where}\quad \xi_c = \{i_{(a,c)} | a\in\partial c\}
    \end{gathered}
    \label{backgr:BP:TN:eq:partition_function}
\end{equation}
and the shorthand $\cntr$ denotes summation over all shared indices. The multi-index $\xi_c$ groups the indices connected to node $c$. The set $\partial c$ contains the nodes adjacent to node $c$.

\begin{figure}
    \begin{center}
        \begin{tikzpicture}
            \input{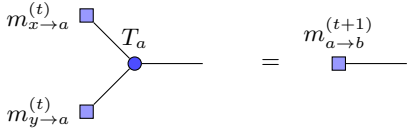}
        \end{tikzpicture}
    \end{center}
    \caption{The message update -- Eq.~\eqref{backgr:BP:TN:eq:msg_update} -- contracts all but one incoming message into the respective tensor. Let $\partial a=\{x,y,b\}$. Then the message $m_{a\rightarrow b}^{(t+1)}$ is obtained by evaluating $\cntr\left(T_am_{x\rightarrow a}^{(t)}m_{y\rightarrow a}^{(t)}\right)$, i.e., by contracting $T_a$ with all incoming messages except the one from edge $(a,b)$.}
    \label{backgr:BP:TN:fig:msgupdate}
\end{figure}

BP is an approximate contraction scheme that estimates $Z$ via an iterative procedure: the BP-iteration. The centerpieces of the algorithm are so-called messages. In the case of TN as introduced thus far, messages are vectors that are associated with an edge and a direction: For each edge $(a,b)$, there are two messages $m_{a\rightarrow b}$ and $m_{b\rightarrow a}$. Let $M^{(t)}=\{m_{a\rightarrow b}^{(t)}|(a,b)\in E^{\leftrightarrow}\}$ denote the set of all messages on a graph at step $t$.\footnote{We use $E^{\leftrightarrow}$ to denote the set of all edges, along with added direction: $E^{\leftrightarrow} = E\times\{\leftarrow, \rightarrow\}$.} The BP iteration is a prescription for the iterative update of $M^{(t)}$ until convergence. Messages may be updated by ``passing them through tensors.'' Consider the message $m_{a\rightarrow b}^{(t)}$: the updated message is the result of contracting the tensor $T_a$ with its incoming messages:
\begin{equation}
    m_{a\rightarrow b}^{(t+1)} = \cntr\left\{T_a\prod_{c\in \partial a\setminus b}m_{c\rightarrow a}^{(t)}\right\}
    \label{backgr:BP:TN:eq:msg_update}
\end{equation}
(omitting additional normalization)~\cite{Alkabetz:2021:TN_BP}. An illustration can be found in Fig.~\ref{backgr:BP:TN:fig:msgupdate}. Equation~\ref{backgr:BP:TN:eq:msg_update} is applied to all messages, until they converge; the corresponding message set $M^{(\infty)} = \lim_{t \to \infty} M^{(t)}$ is referred to as the ``BP fixed point''~\cite{Mezard:2009:InformationPhysicsComputation}. For all numerical simulations in this work, the initial set $M^{(0)}$ comprised random vectors. We demonstrate the effect of the initial set numerically in Sec.~\ref{sec:app:practice:msg}.

\begin{figure*}
    \begin{center}
        \begin{tikzpicture}
            \input{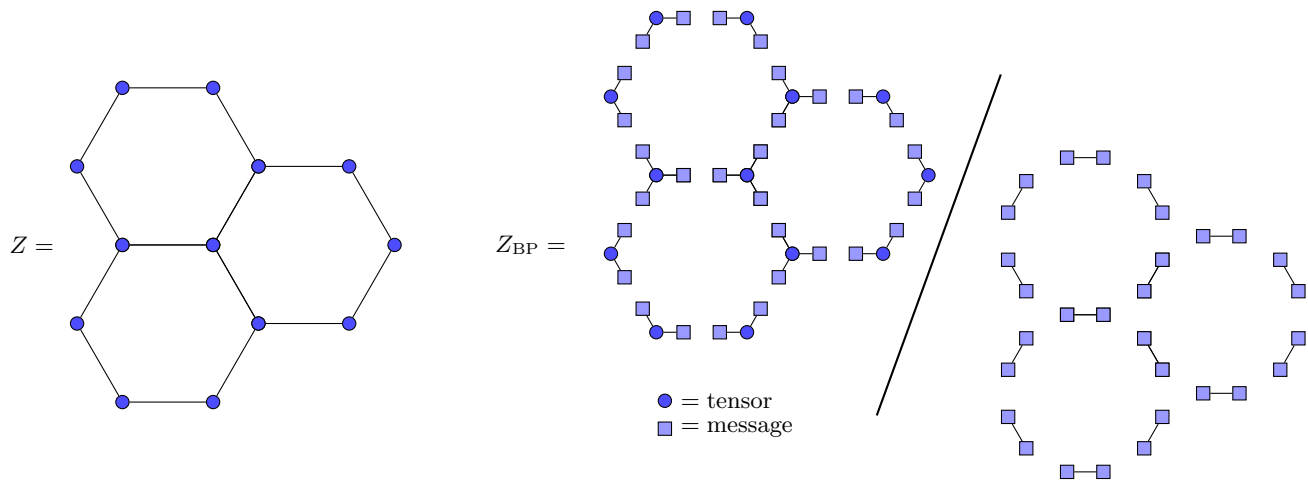}
        \end{tikzpicture}
    \end{center}
    \caption{The contraction value $Z$ of a TN can be obtained using BP by absorbing the messages into the tensors, and dividing by the message normalizations. This example depicts $Z_\text{BP}$ (Eq.~\eqref{backgr:BP:TN:eq:BP_cntr}) for a TN consisting of three hexagonal cells.}
    \label{backgr:BP:TN:fig:hex32_cntr}
\end{figure*}

The estimate $Z_\text{BP}$ of the true contraction value $Z$ is calculated by contracting all fixed-point messages with the tensor they are inbound to. Letting
\begin{equation}
    Z_{a} = \cntr\left\{T_a\prod_{c\in\partial a}m_{c\rightarrow a}\right\},
    \label{backgr:BP:TN:eq:BP_cntr:at_node}
\end{equation}
the estimate becomes
\begin{equation}
    Z_\text{BP} = \prod_{a\in V}Z_a \: \bigg/ \: \prod_{(a,b)\in E}\langle m_{a\rightarrow b},m_{b\rightarrow a}\rangle,
    \label{backgr:BP:TN:eq:BP_cntr}
\end{equation}
where $\langle m_{a\rightarrow b},m_{b\rightarrow a}\rangle$ is the inner product of the two messages on one edge\footnote{Eq.~\eqref{backgr:BP:TN:eq:BP_cntr} stems from the origins of BP in statistical inference. Probability distributions resembling TN are referred to as ``graphical models''; in this picture, the contraction of the TN amounts to finding the distribution's partition function~\cite{Koller:2009:ProbabilisticGraphicalModels, McEliece:1998Tubocodes_and_BP, Pearl:1988:ProbabilisticReasoning}. Recognizing that $Z=\text{exp}(S)$~\cite{Begusic:2024:FastConvergedClassicSim}, with $S$ being the free entropy, Ref.~\cite{Mezard:2009:InformationPhysicsComputation} shows that $S$ may be approximated in terms of the fixed-point messages by the ``Bethe free entropy'' $S_\text{Bethe}$. Evaluating $Z=\text{exp}(S_\text{Bethe})$ yields Eq.~\eqref{backgr:BP:TN:eq:BP_cntr}.} \cite{Begusic:2024:FastConvergedClassicSim}. An illustration can be found in Fig.~\ref{backgr:BP:TN:fig:hex32_cntr}. We will occasionally use the subscript $_\text{BP}$ to denote that a tensor network has been contracted using BP.

A few observations on the procedure thus far:
\begin{itemize}
    \item BP contraction eliminates the necessity for contraction trees, making it both computationally more lightweight and easier to implement.
    \item The intermediate quantities of BP, as introduced thus far, are simple vectors. This makes BP memory-efficient, as a significant bottleneck for the exact contraction of TNs is memory cost~\cite{Gray:2021:HyperContraction, Geiger:2025:OptimizingPartitioning, Cociorva:2002:SpaceTimeTradeOff}.
    \item $Z_\text{BP}$ is exact if the graph $(V,E)$ is a tree. In this case -- omitting normalization -- the message $m_{a\rightarrow b}$ represents the exact contraction of the branch of the tree that is connected to $a$. It follows that $Z_a=\langle m_{a\rightarrow b},m_{b\rightarrow a}\rangle=Z$ for all nodes $a$ and all edges $(a,b)$, and Eq.~\eqref{backgr:BP:TN:eq:BP_cntr} simplifies to $Z_\text{BP}=Z^{|V|}Z^{1-|V|}=Z$.
    \item While BP is guaranteed to converge on a tree, convergence in the presence of loops in the graph cannot be guaranteed. Moreover, if the graph has loops, repeated application of Eq.~\eqref{backgr:BP:TN:eq:msg_update} implies that messages circulate within loops, creating feedback that can cause them to diverge. To combat this, we normalize messages so that they sum to unity.\footnote{The choice of normalization procedure does not affect the BP contraction estimate $Z_\text{BP}$, as message normalization factors cancel out in Eq.~\eqref{backgr:BP:TN:eq:BP_cntr}. Nevertheless, Sec.~\ref{sec:BPDMRG:alg:nondef} will see us use a specific message normalization within the construction of the generalized eigenvalue problems of BP-DMRG.}
\end{itemize}

The practical implementation of BP offers much flexibility with respect to the scheduling of Eq.~\eqref{backgr:BP:TN:eq:msg_update}. Different schedules have been proposed with the goal of, e.g., faster convergence or improved numerical stability. Message updates may be scheduled to pass messages along spanning trees of the graph $(V, E)$~\cite{Wainwright:2003:TreeReParam}, or based on the minimization of the change in messages from one iteration to the next~\cite{Elidan:2006:ResidualBeliefPropagation}. This work employs the most common -- and simplest -- schedule, called ``flooding schedule'': The simultaneous application of Eq.~\eqref{backgr:BP:TN:eq:msg_update} across the entire graph~\cite{Alkabetz:2021:TN_BP, Guo:2023:BlockBP, Mezard:2009:InformationPhysicsComputation, Cao:2017:DEFG}.

We should stress again that, in the case of arbitrary graphs, Eq.~\eqref{backgr:BP:TN:eq:BP_cntr} represents an approximation of the true contraction value. In this work, we will identify two sources of inaccuracy within BP. First, BP contraction accuracy depends on the amount of correlation between distant parts of the TN, which will be the topic of Sec.~\ref{sec:physBP}.\footnote{The role of ``correlations'' becomes clearer in Fig.~\ref{physBP:TNS:fig:BP_norm_rel_err}, where we show that (approximate) product states allow accurate BP contraction of e.g. state norms $\braket{\psi|\psi}$. States that are farther away from product states allow less accurate contraction.} Furthermore, BP contraction accuracy depends on the loops in the graph $(V, E)$. We present numerical results for BP-contraction of TN in Sec.~\ref{sec:app:practice:cntr}, finding that BP contraction accuracy improves exponentially with growing loop length.\footnote{The loops within a TN, and the sources of error they represent in the context of BP, point the way to the systematic improvement of BP. The authors of Ref.~\cite{Evenbly:2026:LoopSeriesExpansionPublished} show how loops within a TN can be systematically evaluated, resulting in a controlled series of corrections to the BP contraction value $Z_\text{BP}$.}

\subsubsection{Computing Expectation Values}
\label{sec:backgr:BP:QMBP}

TN have first proven useful for quantum physics in the form of tensor network states (TNS) and tensor network operators (TNO). We shall introduce TNS and TNOs first, before moving on to the calculation of quantities of interest from those.

\vspace{\baselineskip}

\begin{figure}
    \begin{center}
        \begin{tikzpicture}
            \input{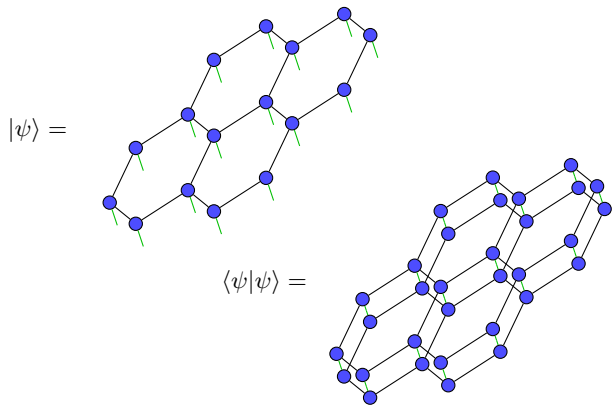}
        \end{tikzpicture}
    \end{center}
    \caption{The norm $\braket{\psi|\psi}$ can be expressed as a tensor network, given a TNS representation of $\ket{\psi}$. The conjugated state $\bra{\psi}$, in TNS format, is represented by complex conjugates $T^\dagger$ of all TNS-tensors $T$ of the state $\ket{\psi}$. Mathematically, the inner product $\braket{\psi|\psi}$ is a TN that is obtained by connecting the physical legs of $T_a$ and $T_a^\dagger$ for all nodes $a\in V$.}
    \label{backgr:BP:QMBP:fig:hex22_norm}
\end{figure}

TNS are a class of Ans{\"a}tze for wavefunctions, where one writes the amplitudes as TN contractions:
\begin{equation}
    \ket{\psi} = \sum_{\overline{x}}\psi_{\overline{x}}\ket{\overline{x}}, \quad\text{where}\quad \psi_{\overline{x}} = \cntr\left\{\prod_{a\in V}T_a^{[x_a]}\right\}.
    \label{backgr:BP:QMBP:eq:TNS}
\end{equation}
The wavefunction $\ket{\psi}$ of an $N$-particle system can thus be depicted as a TN with $N$ open indices (often termed ``physical legs''). The index $\overline{x} = \{x_a|a\in V\}$ runs over the basis of the Hilbert space. Most often, this TN contains one tensor for each of the $N$ sites. Tensors share legs (``virtual legs'') whose size is termed the ``bond dimension'' $\chi$. (See Fig.~\ref{backgr:BP:QMBP:fig:hex22_norm} for an illustration.) A TNO represents an analogous way of writing the matrix elements of an operator $O$ as TNs, such that an entire operator may be represented as a TN with physical legs:
\begin{equation}    
    \begin{gathered}
        O = \sum_{\overline{x},\overline{y}} O_{\overline{x},\overline{y}}\ket{\overline{x}}\bra{\overline{y}}, \quad\text{where} \\
        O_{\overline{x},\overline{y}} = \cntr\left\{\prod_{a\in V}W_a^{[x_a,y_a]}\right\}.
    \end{gathered}
    \label{backgr:BP:QMBP:eq:TNO}
\end{equation}
Just like the $T_a$ in Eq.~\eqref{backgr:BP:QMBP:eq:TNS}, within this work, the $W_a$ are tensors that are associated with lattice sites of the system. The tensors $W_a$ possess two physical legs each, and are connected to each other with edges whose size, too, is referred to as ``bond dimension''. Constructing a TNO is possible for arbitrary operators, but may incur large bond dimensions depending on the choice of the topology and the operator itself~\cite{Schollwoeck:2011:DMRG}. If, on the contrary, the operator $O$ has a small support or consists of short operator strings, a TNO with small and constant bond dimensions can be constructed~\cite{Bridgeman:2017:InterpretativeDance, Hubig:2017:GenericMPO, Paeckel:2017:MPOConstruction, Ren:2020:Bipartite, Cakir:2025:SymbolicTTNO}. Here, we rely on the particle-decay construction~\cite{Bridgeman:2017:InterpretativeDance}, which we introduce in Sec.~\ref{sec:app:MPO}.

The most prominent applications of TNS are matrix-product states (MPS) on one-dimensional geometries, and pair-entangled product states (PEPS) on two-dimensional lattices~\cite{Orus:2014:MPS_PEPS:intro}. Note that these two examples are limited to specifically structured lattices (for example, the one-dimensional line in the case of MPS). This limitation is not inherent to TNS, however: One may imagine an arbitrary graph and place tensors $T_a$ on the nodes, thereby creating a TNS on that graph. Tensors then share edges according to the edges of the graph.\footnote{Note that a TNS or a TNO must not have an edge for each edge of the underlying lattice; the possibilities for connecting TNS tensors are manifold. Consider, e.g., a rectangular grid of spins which is much shorter on one side than on the other. A TNS on such a system may be chosen as PEPS, or one may construct an MPS that ``winds'' through the graph~\cite{Stoudenmire:2012:TwoDimDMRG} (see Fig.~\ref{backgr:DMRG:MPS:fig:2d_mps_snake}). The former allows us to exploit the benefits that low entanglement implies for bond dimensions, while the latter provides greater numerical stability in DMRG.}

\vspace{\baselineskip}

The utility of such a representation comes from the fact that the ground states of local, gapped Hamiltonians may be efficiently represented as TNS, due to the ``area law'': The entanglement of ground states of local, gapped Hamiltonians is bound by the size of the boundary of the subsystem under consideration~\cite{Eisert:2010:AreaLaws, Schollwoeck:2011:DMRG}. Noting that the difficulties in simulating quantum systems can often be traced back to large amounts of entanglement within them~\cite{Orus:2014:MPS_PEPS:intro}, the area law implies that the virtual bond dimensions of the tensors that constitute a TNS can be significantly smaller compared to a generic state, thereby exploiting structure that is hidden within quantum states.

\vspace{\baselineskip}

Assume then that a TNS is at hand, and one wishes to calculate quantities of interest from it. Consider, for example, the norm $\braket{\psi|\psi}$ of the state $\ket{\psi}$; with the state being in TNS format, this becomes a TN, by connecting the physical legs appropriately (see Fig.~\ref{backgr:BP:QMBP:fig:hex22_norm} for an illustration).

The calculation of $\braket{\psi|\psi}$, in the context of arbitrary geometries, thus inherits the complexities associated with TN contraction: contraction tree optimization and memory costs. At this point, BP enters the picture.

\vspace{\baselineskip}

As Section~\ref{sec:backgr:BP:TN} outlines, BP can be used for TN contraction; one may thus employ it to calculate an estimate $\braket{\psi|\psi}_\text{BP}$. This is done by passing messages not between individual tensors in the network (as this would include, e.g., $T_a$ and $T_a^\dagger$ at the same physical site), but between lattice sites. We may re-write Eq.~\eqref{backgr:BP:TN:eq:msg_update} to include the TNS tensor $T$ and its adjoint, arriving at
\begin{equation}
    m_{a\rightarrow b}^{(t+1)} = \cntr\left\{T_aT_a^\dagger\prod_{c\in \partial a\setminus b}m_{c\rightarrow a}^{(t)}\right\}.
    \label{backgr:BP:QMBP:eq:msg_update}
\end{equation}
Although the formulation is nearly equivalent, the meaning of the involved quantities changes. First, the subscripts $a,b,c\in V$ do not refer to an individual tensor anymore, but to a physical site within the lattice. Moreover, Eq.~\eqref{backgr:BP:QMBP:eq:msg_update} implies that a message becomes a matrix: $m_{a\rightarrow b}\in\mathbb{C}^{\chi\times\chi}$.

Equations \eqref{backgr:BP:TN:eq:BP_cntr:at_node} and \eqref{backgr:BP:TN:eq:BP_cntr} may be modified accordingly, adapting BP to the contraction of $\braket{\psi|\psi}$. This technique is known as ``lazy BP'', since the contraction $\cntr(T_a T_a^\dagger)$ does not need to be performed, reducing the memory footprint~\cite{Begusic:2024:FastConvergedClassicSim}.

Note further that BP can, in this way, be executed not only on norms $\braket{\psi|\psi}$, but also on expectation values $\braket{\psi|O|\psi}$ and, furthermore, on objects like $\braket{\psi|O^\dagger O|\psi}$, where $O$ is a TNO.\footnote{The interpretation of BP messages in the case of expectation values relates to the blocks of a DMRG calculation. Messages on an expectation value $\braket{\psi|O|\psi}$, where the lattice is tree-shaped, are exact contractions of branches of the tree; thus, BP generates the blocks of DMRG that contain the environment. We go into more detail on this in Sec.~\ref{sec:BPDMRG:alg:nondef}.} Analogous to norms, one connects physical legs accordingly, and passes messages between physical sites of the TN. Where messages on $\braket{\psi|\psi}$ are two-legged tensors, messages on $\braket{\psi|O|\psi}$ are three-legged, and so forth.

\vspace{\baselineskip}

Finally, for BP to be a means for accurate contraction, the respective TN must be \textit{sign-preserving}, by which we mean that the BP iteration converges to fixed point messages $M^{(\infty)}$ from which the correct sign of $Z_\text{BP}$ can be extracted. BP inherits this necessity from its origins in statistical inference. The derivation of Eq.~\eqref{backgr:BP:TN:eq:msg_update} involves mapping the TN in question to a probability distribution, which is written as a ``normal factor graph''~\cite{Forney:2001:NormalCodes}. Probability distributions being an intermediate step towards TN contraction means that the TN must maintain an appropriate notion of positivity, which -- in the context of simple TN -- can be satisfied by restricting the value range of tensors to positive numbers.

The meaning of this ``appropriate notion of positivity'' changes in the context of TNS and expectation values, as site tensors $T_a\in\mathbb{C}^{\chi^{|\partial a|}\times D}$ of quantum states do not consist of positive numbers in the general case. However, the quantity $\braket{\psi|\psi}$ is strictly non-negative. The BP iteration we have presented in Sec.~\ref{sec:backgr:BP:QMBP} is designed to ensure that $\braket{\psi|\psi}_\text{BP}\geq 0$ as well: If, at any step $t$, all messages $m_{a\rightarrow b}^{(t)}$ are positive semi-definite (PSD), Eq.~\eqref{backgr:BP:QMBP:eq:msg_update} results in PSD messages $m_{a\rightarrow b}^{(t+1)}$.\footnote{Ensuring that messages remain positive-semidefinite is the reason why messages are passed between lattice sites instead of tensors on $\braket{\psi|\psi}$. The BP iteration from Sec.~\ref{sec:backgr:BP:TN} needs to be modified, since it is not sign-preserving when applied to $\braket{\psi|\psi}$. Formally, Eq.~\eqref{backgr:BP:QMBP:eq:msg_update} executes BP on so-called ``double-edged factor graphs''~\cite{Cao:2017:DEFG}, implementing a sign-preserving message update.} When the BP fixed point $M^{(\infty)}$ consists of PSD-messages, $\braket{\psi|\psi}_\text{BP}$ is guaranteed to be non-negative as well. On $\braket{\psi|H|\psi}$, too, the BP iteration is only guaranteed to converge if the message update protects the definiteness of the messages; this is the case if $H$ is positive- or negative-semidefinite. We discuss sign-preserving TN, and more importantly the effects of a lack thereof, in Sec.~\ref{sec:app:practice:nondef}.

\subsection{DMRG}
\label{sec:backgr:DMRG}

The DMRG algorithm aims to find the ground states of many-particle Hamiltonians. It was developed independently from the tensor network formulation of QMBP, and its success owes to the proper incorporation of boundary conditions between a small part of the system and the larger whole, enabling the treatment of systems comprising many particles while using only modest computational resources~\cite{Verstraete:2023:30yearsDMRG, White:1992:DMRG, White:1992:QuantumDMRG}. Only later was it discovered that DMRG is a prescription for iteratively optimizing MPS~\cite{Dukelsky:1998:DMRGinMPS}. This discovery opened the floodgates to many improvements and variants of the original DMRG; the rich picture of local, gapped Hamiltonians, whose ground states follow area laws and may be efficiently represented using TNS, followed suit~\cite{Schollwoeck:2011:DMRG, Verstraete:2004:QMBP_renormalization, Verstraete:2004:QuantumInfoDMRG}.

\vspace{\baselineskip}

For the purposes of this work, we will consider DMRG as an iterative method for (greedy) optimization of TNS. We will lay the groundwork by introducing DMRG on tree tensor network states (TTNS), and expand on it with the construction of BP-DMRG.\footnote{In what follows, we draw heavily from Refs.~\cite{Schollwoeck:2011:DMRG, Bridgeman:2017:InterpretativeDance}.}

Consider thus the problem of ground state search, i.e., the task of finding
\begin{equation}
    \ket{E_0} = \argmin_{\ket{\psi}\in\mathcal{H}} \frac{\braket{\psi|H|\psi}}{\braket{\psi|\psi}},
    \label{backgr:DMRG:eq:rayleigh_coeff}
\end{equation}
where the state $\ket{\psi}$ is represented by a TTNS. In what follows, we assume that the Hamiltonian $H$ is represented as a tree tensor network operator (TTNO), with site tensors denoted by $W_a$. The only degrees of freedom in a TTNS are the site tensors $T_a$, so an iterative solver of Eq.~\eqref{backgr:DMRG:eq:rayleigh_coeff} finds
\begin{equation}
    T_a^\text{(next)} = \argmin_{T_a} \frac{\braket{\psi(T_a)|H|\psi(T_a)}}{\braket{\psi(T_a)|\psi(T_a)}}
    \label{backgr:DMRG:eq:rayleigh_coeff:MPS}
\end{equation}
at every physical site $a$.

The first step towards solving Eq.~\eqref{backgr:DMRG:eq:rayleigh_coeff:MPS} efficiently is recognizing that solutions may be obtained by extremizing $\braket{\psi(T_a)|H|\psi(T_a)}-\lambda\braket{\psi(T_a)|\psi(T_a)}$. Evaluating both terms gives
\begin{widetext}
    \begin{gather}
        \braket{\psi(T_a)|H|\psi(T_a)} = \cntr\left\{\prod_{b\in V} T_b^\dagger W_bT_b\right\} = \cntr\Bigg\{T_a^\dagger W_a T_a \underbrace{\prod_{b\neq a} T_b^\dagger W_b T_b}_{\equiv E} \Bigg\} = \cntr \left( T_a^\dagger W_aT_a E \right) = \braket{T_a|H_a|T_a},
        \label{backgr:DMRG:eq:local_hamiltonian} \\
        \braket{\psi(T_a)|\psi(T_a)} = \cntr\Bigg\{T_a^\dagger T_a \underbrace{\prod_{b \neq a} T_b T_b^\dagger}_{\equiv N_a} \Bigg\} = \braket{T_a|N_a|T_a}.
        \label{backgr:dmrg:eq:local_environment}
    \end{gather}
\end{widetext}

\begin{figure}
    \begin{center}
        \begin{tikzpicture}
            \input{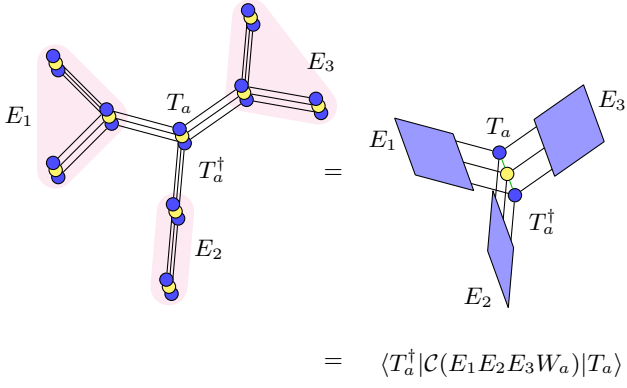}
        \end{tikzpicture}
    \end{center}
    \caption{A local Hamiltonian is the result of contracting all of $\braket{\psi|H|\psi}$ except the respective site tensors, yielding $\braket{T_a|H_a|T_a}$ at site $a$. This is the graphical depiction of Eq.~\eqref{backgr:DMRG:eq:local_hamiltonian}, with TTNS tensors drawn in blue and TTNO tensors drawn in yellow.}
    \label{backgr:DMRG:fig:local_hamiltonian}
\end{figure}

Eq.~\eqref{backgr:DMRG:eq:local_hamiltonian} amounts to contracting the ``sandwich'' $\braket{\psi|H|\psi}$ around site $a$, where the tensor $E$ contains all of $\braket{\psi|H|\psi}$ except site $a$. By writing $H_a = \cntr(W_a E)$ and vectorizing $T_a\mapsto\ket{T_a}$, Eq.~\eqref{backgr:DMRG:eq:local_hamiltonian} becomes the matrix-vector product $\braket{T_a|H_a|T_a}$. Proceeding analogously with Eq.~\eqref{backgr:dmrg:eq:local_environment} gives $\braket{\psi(T_a)|\psi(T_a)}=\braket{T_a|N_a|T_a}$. We refer to $H_a$ and $N_a$ as the ``local Hamiltonian'' and ``local environment'', respectively. Solving Eq.~\eqref{backgr:DMRG:eq:rayleigh_coeff:MPS} is thus equivalent to solving a generalized eigenvalue problem:
\begin{equation}
    H_a\ket{T_a} = \lambda N_a\ket{T_a}.
    \label{backgr:DMRG:eq:gen_eigval}
\end{equation}
The procedure for constructing the local Hamiltonian in the case of $\ket{\psi}$ being a TTNS is shown in Fig.~\ref{backgr:DMRG:fig:local_hamiltonian}. In this example, the environment $E$ consists of three disjoint branches which can be contracted separately. The local Hamiltonian then becomes the contraction of all environments with the TTNO tensor $W_a$.

\begin{figure}
    \begin{center}
        \begin{tikzpicture}
            \input{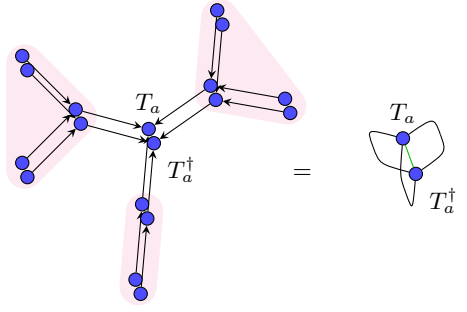}
        \end{tikzpicture}
    \end{center}
    \caption{When a TTNS is orthogonalized around a site $a$ (indicated by drawing edges as arrows pointing towards the orthogonality center), the contraction of $\braket{\psi|\psi}$ around $a$ (Eq.~\eqref{backgr:dmrg:eq:local_environment}) simplifies, since all branches of the TTNS adjacent to $a$ (shown as shaded in magenta) simplify to the identity. This ensures the numerical stability of DMRG on TTNS.}
    \label{backgr:DMRG:fig:local_environment}
\end{figure}

Note, however, that Eq.~\eqref{backgr:DMRG:eq:gen_eigval} represents a \textit{generalized} eigenvalue problem, which can be numerically unstable, if the condition number of $N_a$ is large~\cite{Schollwoeck:2011:DMRG, Bridgeman:2017:InterpretativeDance}. The canonical method for circumventing this problem is to keep the TTNS in a so-called ``canonical form''. By exploiting the gauge freedom along the virtual edges between tensors, the TTNS tensors may be kept in a form where the contraction of a branch of the tree with its adjoint equals the identity. This can be done by choosing $a$ as the root node of the tree. The state is then gauged through successive QR-decompositions, starting at the leaves~\cite{Evenbly:2022:PracticalGuideTN}. We will refer to the site $a$ as ``orthogonality center'' (see Fig.~\ref{backgr:DMRG:fig:local_environment}). Eq.~\eqref{backgr:DMRG:eq:gen_eigval} now becomes a -- numerically stable -- standard eigenvalue problem.

\vspace{\baselineskip}

We thus arrive at a variational, greedy method in the space of fixed-bond-dimension TTNS. In a preparatory step, the TTNS is gauged around the site to be optimized. After construction of the local Hamiltonian by partial contraction of $\braket{\psi|H|\psi}$ and solving the eigenvalue problem, a new site tensor $T_a^{\text{(new)}}$ is found and inserted into the TTNS. The orthogonality center then moves to the next site, and the procedure is repeated.

In practice, given a physical system, both the choice of the tree that defines the TTNS topology and the order in which sites are optimized influence the algorithm's computational cost. The tree itself should be chosen to minimize entanglement between sites that are distant in the tree; the traversal order should be chosen such that maintaining the canonical form of the TTNS is computationally cheap. A depth-first traversal order satisfies the latter requirement~\cite{Nakatani:2013:TreeDMRG}. Once a traversal order is chosen, environments from $\braket{\psi|H\psi}$ may be cached, further accelerating numerical implementations.

During runs of DMRG, the energy expectation value is guaranteed to monotonically decrease~\cite{Bridgeman:2017:InterpretativeDance}, however, DMRG might fall into sub-optimal local minima within the global energy landscape of the Hamiltonian~\cite{Schollwoeck:2011:DMRG}.

\vspace{\baselineskip}

\begin{figure}
    \begin{center}
        \begin{tikzpicture}
            \input{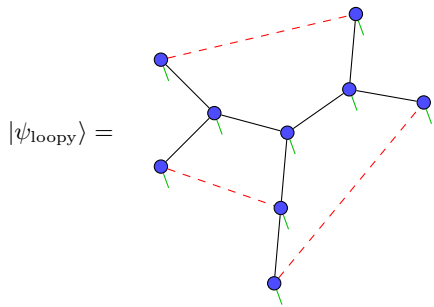}
        \end{tikzpicture}
    \end{center}
    \caption{TTNS are Ansatz states on which DMRG can be executed in a computationally efficient manner. As soon as loops are introduced into the topology -- here indicated by dashed edges in red -- canonical forms are rarely available, making DMRG challenging computationally.}
    \label{backgr:DMRG:fig:loopyTTNS}
\end{figure}

The development of BP-DMRG is motivated by the difficulties of DMRG in the presence of loops or on irregular lattices. Fig.~\ref{backgr:DMRG:fig:loopyTTNS} shows a TTNS with three additional edges, which create loops in the underlying graph. Given a system in which interactions between sites form loops, an Ansatz that shares the system's topology is likely to minimize entanglement between sites. This is one of the requirements for computational efficiency we laid out earlier; however, since the TNS contains loops, canonical forms are no longer readily available, and a natural ordering for optimizing the nodes is lacking.

While canonical forms in loopy topologies do exist, they are often approximate~\cite{Gray:2024:HyperApproxContraction}, or come with limitations w.r.t.\ their variational power~\cite{Zaletel:2020:isoPEPS, Sappler:2025:DiagonalIsoTPSPublished}. If the lack of canonical forms were to be mitigated by mapping the system's interactions to a tree TTNO and choosing a corresponding Ansatz for $\ket{\psi}$, it is not obvious which tree is optimal~\cite{Cakir:2025:SymbolicTTNO}.

An important architectural choice in implementing loopy DMRG is thus the topology of the Ansatz state. In order to motivate the development of BP-DMRG, we shall discuss two possible choices: tree topologies (TTNS-Ans\"{a}tze), or loopy topologies (general TNS).

\subsubsection{Tree Topologies}
\label{sec:backgr:DMRG:TTNS}

On the one hand, one might choose a TTNS as an Ansatz wavefunction. This is advantageous since the DMRG algorithm outlined above can be used as-is, requiring no modification, and offers guaranteed convergence and numerical stability.

The difficulty is that the physical system must be mapped to a tree graph; this may necessitate large bond dimensions in both the TTNS and the TTNO.

Technically, within TTNOs, operator chains grow longer when physical sites cease to be adjacent in the Ansatz state.\footnote{This can be understood through the ``particle decay construction'' of Tensor Product Operators. It prescribes constructing TNOs as operator chains by weaving them through TNO site tensors to minimize the required bond dimensions. We discuss the construction of TNOs in more detail in Sec.~\ref{sec:app:MPO}.}

Physically, within TTNS, entanglement between sites that are not adjacent in the Ansatz leads to large bond dimensions along the section of the TTNS connecting the two sites. In a naive example, a one-dimensional Ansatz on a system with loopy interactions may lead to exponentially increasing bond dimensions in the TTNS~\cite{Stoudenmire:2012:TwoDimDMRG}.

\begin{figure}
    \centering
    \begin{tikzpicture}
            \pgfmathsetmacro{\k}{.5}

            \foreach \i in {0,1,2,3,4,5,6,7,8,9,10,11,12,13,14,15}{
                \foreach \j in {0,1,2}{
                    \node (t-\i-\j) at ($(\k*\i,\k*\j)$) [tpsnode] {};
                }
            }
            \foreach \i in {0,1,2,3,4,5,6,7,8,9,10,11,12,13,14,15}{
                \foreach \j/\l in {0/1,1/2}{
                    \draw [thick] (t-\i-\j) -- (t-\i-\l);
                }
            }
            \foreach \i/\l in {0/1,1/2,2/3,3/4,4/5,5/6,6/7,7/8,8/9,9/10,10/11,11/12,12/13,13/14,14/15}{
                \foreach \j in {0,1,2}{
                    \draw [densely dotted] (t-\i-\j) -- (t-\l-\j);
                }
            }
            \foreach \i/\l in {1/2,3/4,5/6,7/8,9/10,11/12,13/14}{
                \draw [thick] (t-\i-0) -- (t-\l-0);
            }
            \foreach \i/\l in {0/1,2/3,4/5,6/7,8/9,10/11,12/13,14/15}{
                \draw [thick] (t-\i-2) -- (t-\l-2);
            }
        \end{tikzpicture}
    \caption{MPS may be employed to solve ground state problems in geometries other than the line. In the example of a narrow, rectangular strip of spins, this can be achieved by letting the MPS ``snake'' through the system. (In this example, all edges, whether dotted or solid, symbolize coupling between spins; only solid edges are edges within the MPS.) This approach inherits the advantages of 1D-DMRG: numerical stability and ease of implementation, but comes at the cost of increased bond dimensions. Figure inspired by Ref.~\cite{Stoudenmire:2012:TwoDimDMRG}.}
    \label{backgr:DMRG:MPS:fig:2d_mps_snake}
\end{figure}
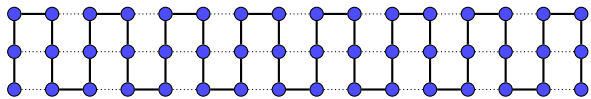

An example for mapping a system with pairwise interaction on a two-dimensional lattice to a one-dimensional Ansatz is shown in Fig.~\ref{backgr:DMRG:MPS:fig:2d_mps_snake}.

\subsubsection{Topologies with Loops}
\label{sec:backgr:DMRG:TNS}

The other option is to choose an Ansatz that reflects the system's topology. This is advantageous with respect to the required bond dimensions in both the TNS and the TNO.


If the graph of the TNS mirrors the system's topology, the cumulative bond dimensions across boundaries between bipartitions may follow an area law, mimicking the entanglement of the ground state of a local Hamiltonian. This removes the exponential scaling of bond dimensions that MPS or TTNS may exhibit.

\vspace{\baselineskip}

Choosing either tree- or loopy topologies comes with other difficulties, however: contraction and numerical stability.

While the canonical form of TTNS, as we have seen, removes much of the numerical workload associated with contraction of $\braket{\psi|\psi}$, canonical forms of TNS are not readily available in higher dimensions.\footnote{A notable exception are the ``isoPEPS'', introduced in Refs.~\cite{Sappler:2025:DiagonalIsoTPSPublished, Zaletel:2020:isoPEPS}. The authors define a canonical form for two-dimensional TNS that admits trivial contraction into an orthogonality center. This, however, limits the variational power of such states.} Instead, adapting Eqs.~\eqref{backgr:DMRG:eq:local_hamiltonian} and \eqref{backgr:dmrg:eq:local_environment} amounts to repeated contractions of a large TN without the help of canonical forms.

Furthermore, the canonical form of TTNS ensures that the generalized eigenvalue problem within DMRG is well-conditioned.

\vspace{\baselineskip}

Having introduced both the advantages of the DMRG algorithm and the difficulties one faces when adapting it to less-structured, loopy topologies, we will propose a novel approach to addressing these difficulties in the next section: combining DMRG with belief propagation.

\section{BP-DMRG}
\label{sec:BPDMRG}

The insight that underpins the development of BP-DMRG is that a DMRG-like algorithm, in essence, consists of the repeated evaluation of Eq.~\eqref{backgr:DMRG:eq:rayleigh_coeff:MPS}. BP-DMRG, through this lens, proposes to incorporate belief propagation into DMRG as a subroutine for rapid contraction.

This section presents the BP-DMRG algorithm and evaluates its performance for ground state search. We are inspired by Ref.~\cite{Alkabetz:2021:TN_BP}, where belief propagation is used to calculate reduced density matrices, suggesting that messages can be seen as contractions of parts of the TN even in the presence of loops. The idea here is thus to use belief propagation during the construction of the generalized eigenvalue problem. We provide a brief overview of the key characteristics of BP-DMRG:
\begin{itemize}
    \item Ans\"{a}tze on loopy topologies instead of tree topologies are used, limiting the growth of bond dimensions and keeping with our goal of exploring lattice-agnostic solutions to the ground state problem. The absence of canonical forms and the need for more efficient contraction techniques motivate the usage of BP.
    \item Canonical forms are approximated by a tree-equivalent gauge, mitigating the ill-conditioning of generalized eigenvalue problems.
    \item BP enables rapid, approximate contraction of $\braket{\psi|H|\psi}$ and $\braket{\psi|\psi}$, and is used to construct local Hamiltonians and environments.
\end{itemize}

\subsection{The Algorithm}
\label{sec:BPDMRG:alg}

We will devote two subsections to different aspects of the BP-DMRG algorithm, each building on DMRG itself and the preceding content: how to use BP for $\braket{\psi|H|\psi}$, and the numerical stability of the generalized eigenvalue problems.

\subsubsection{BP for \texorpdfstring{$\braket{\psi|H|\psi}$}{non-definite expectation values}}
\label{sec:BPDMRG:alg:nondef}

In order to adapt DMRG to arbitrary lattices, a method to efficiently construct the local Hamiltonians $H_a$ and local environments $N_a$ is needed. This section explains how messages can be employed for this task and how messages on $\braket{\psi|H|\psi}$ can be obtained in the first place.

\vspace{\baselineskip}

Since messages (approximately) represent contractions of the respective TN around the nodes to which they are inbound, they may be used to calculate the environments that enter into $H_a$ and $N_a$.

Consider first the local Hamiltonian $H_a$. Once the BP fixed point for $\braket{\psi|H|\psi}$ has been found, and messages have been normalized to the contraction value $\braket{\psi|H|\psi}_\text{BP}$,\footnote{Normalization of messages to contraction value implies that the contraction value $Z_a$ of a site tensor $T_a$ and its incoming messages is equal to the BP contraction value of the complete network: $Z_a = Z_\text{BP}$ for all $a\in V$ (refer back to Eq.~\eqref{backgr:BP:TN:eq:BP_cntr:at_node}). Once BP fixed point messages have been obtained and $Z_\text{BP}$ is available, this is done by setting $m_{b\rightarrow a} \mapsto \frac{Z_\text{BP}}{Z_a^{1/|\partial a|}}m_{b\rightarrow a}$.} we approximate the environment block $E$ (Eq.~\eqref{backgr:DMRG:eq:local_hamiltonian}) by the product of inbound messages. The local Hamiltonian thus becomes
\begin{equation}
    H_a = \cntr\left(W_a\prod_{b\in\partial a}m_{b\rightarrow a}\right).
    \label{BPDMRG:alg:nondef:eq:local_Hamiltonian}
\end{equation}
This construction is exact on tree-shaped lattices, where it reproduces the result that we illustrate in Fig.~\ref{backgr:DMRG:fig:local_hamiltonian}. In the presence of loops, Eq.~\eqref{BPDMRG:alg:nondef:eq:local_Hamiltonian} implicitly uses the assumption that the environment block factorizes.

\begin{figure}
\centering
\begin{tikzpicture}
    \pgfmathsetmacro{\k}{.5}
    \pgfmathsetmacro{\d}{3*\k/12}

    \node (W) at (0,0) [tponode, label=225:$W_a$] {};
    \foreach \i/\p in {1/330, 2/50, 3/180}{
        \coordinate (box\i_inner) at ($(W)+(\p:3*\k)$);
        \coordinate (box\i_outer) at ($(box\i_inner)+(\p:\k)$);
        \coordinate (box\i_anchor1) at ($(box\i_inner)+(0,\k)$);
        \coordinate (box\i_corner1) at ($(box\i_anchor1)+(0,\d)+(180+\p:\d)$);
        \coordinate (box\i_anchor2) at ($(box\i_inner)+(0,-\k)$);
        \coordinate (box\i_corner2) at ($(box\i_anchor2)+(0,-\d)+(180+\p:\d)$);
        \coordinate (box\i_anchor3) at ($(box\i_outer)+(0,-\k)$);
        \coordinate (box\i_corner3) at ($(box\i_anchor3)+(0,-\d)+(\p:\d)$);
        \coordinate (box\i_anchor4) at ($(box\i_outer)+(0,\k)$);
        \coordinate (box\i_corner4) at ($(box\i_anchor4)+(0,\d)+(\p:\d)$);

        \draw (W) -- (box\i_inner);
        \draw (box\i_anchor1) -- ($(box\i_anchor1)+(\p+180:2*\k)$);
        \draw (box\i_anchor2) -- ($(box\i_anchor2)+(\p+180:2*\k)$);

        \draw [msgbase, rounded corners=\k] (box\i_corner1) -- (box\i_corner2) -- (box\i_corner3) -- (box\i_corner4) -- cycle;
    }

    \draw [physedge] ($(W)+(0,-\k)$) -- (W) -- ($(W)+(0,\k)$);

    \node (box1_label) at (box1_outer) [label=right:$m_{b\rightarrow a}$] {};
    \node (box2_label) at (box2_outer) [label=right:$m_{c\rightarrow a}$] {};
    \node (box3_label) at (box3_outer) [label=left:$m_{d\rightarrow a}$] {};

    \node (math) at ($(W)+(110:3*\k)$) [label=left:{$H_a=\cntr\big(W_am_{b\rightarrow a}m_{c\rightarrow a}m_{d\rightarrow a}\big)=$}] {};
\end{tikzpicture}
\caption{Eq.~\eqref{BPDMRG:alg:nondef:eq:local_Hamiltonian} describes the calculation of local Hamiltonians at sites with an arbitrary number of neighbors. This figure shows $H_a$, where $|\partial a|=3$. The expression is symbolically similar to $\cntr(E_1 E_2 E_3 W_a)$ from Fig.~\ref{backgr:DMRG:fig:local_hamiltonian}, hinting at the fact that Eq.~\eqref{BPDMRG:alg:nondef:eq:local_Hamiltonian} is equivalent to Eq.~\eqref{backgr:DMRG:eq:local_hamiltonian} on trees.}
\label{BPDMRG:alg:nondef:fig:threeadjH}
\end{figure}
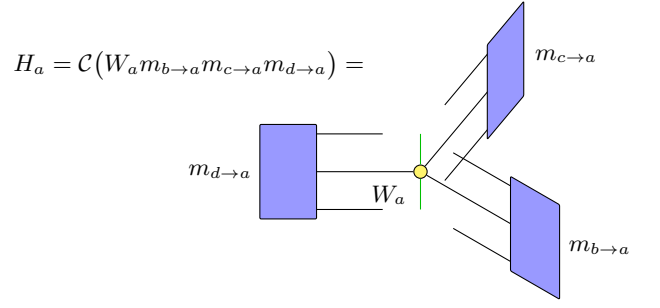

The local Hamiltonian is thus at hand; the approach for the local environment $N_a$ is analogous: by adapting Eq.~\eqref{backgr:dmrg:eq:local_environment}, $N_a$ becomes the outer product of the messages on $\braket{\psi\psi}$. Since $N_a$ also has physical legs, which the messages do not, an identity $I$ with dimension $D\times D$ is added:\footnote{The operators that obtain from Eqs.~\eqref{BPDMRG:alg:nondef:eq:local_Hamiltonian:directsum} and \eqref{BPDMRG:alg:nondef:eq:local_environment} may be highly sparse. Since these quantities are tensor networks, it can be computationally beneficial to store their constituents (messages and TNO site tensors) instead. Calculating, e.g., $N_a\ket{v}$ can then be done in a matrix-free fashion, potentially accelerating the solution of the generalized eigenvalue problem.}
\begin{equation}
    N_a = \cntr\left(I\prod_{b\in\partial a}m_{b\rightarrow a}\right).
    \label{BPDMRG:alg:nondef:eq:local_environment}
\end{equation}

To define the generalized eigenvalue problem, we thus need to execute BP-DMRG on $\braket{\psi|H|\psi}$ and $\braket{\psi\psi}$. At this point, however, another difficulty arises: As we discuss in Sec.~\ref{sec:backgr:BP:QMBP}, BP is not guaranteed to converge to $\braket{\psi|H|\psi}$, if $H$ is neither positive- nor negative-semidefinite. We circumvent this problem by splitting up the TNO $H$ into a positive-semidefinite and a negative-definite part.

Hamiltonians are typically defined (symbolically) as a ``sum of operator strings''. Consider, e.g., the transverse-field Ising model: its operator strings are the pairwise terms $JZ_a\otimes Z_b$, and the external fields $gX_a$. We can now split the operator $H=H^{(+)}+H^{(-)}$ into a PSD- and an NSD-part, such that $\braket{\psi|H|\psi}=\braket{\psi|H^{(+)}|\psi}+\braket{\psi|H^{(-)}|\psi}$. This is done by splitting all operator strings accordingly, for example, for the TFI:
\begin{equation}
    \begin{aligned}
        JZ_a\otimes Z_b =&\phantom{+} \underbrace{J\ket{00}\bra{00}+J\ket{11}\bra{11}}_{\text{PSD}}\\ &+ \underbrace{\left(-J\ket{01}\bra{01}-J\ket{10}\bra{10}\right)}_{\text{NSD}}.
    \end{aligned}
    \label{BPDMRG:alg:nondef:eq:ZZ_split}
\end{equation}
We find that splitting all operator strings and constructing two TNOs, $H^{(+)}$ and $H^{(-)}$ from them, ensures reliable convergence of BP. Note that running BP on $\braket{\psi|H^{(+)}|\psi}$, $\braket{\psi|H^{(-)}|\psi}$, and $\braket{\psi|\psi}$ yields three distinct sets of messages; We will denote messages as $m_{a\rightarrow b}^{(\pm)}$ or $m_{a\rightarrow b}$ according to the TN they belong to.

Executing BP on $\braket{\psi|H^{(\pm)}|\psi}$ and $\braket{\psi|\psi}$, the task when constructing the generalized eigenvalue problem is then the following: construct the local Hamiltonian $H_a$ from the messages $\{m_{b\rightarrow a}^{(\pm)}|b\in\partial a\}$, and construct $N_a$ from $\{m_{b\rightarrow a}|b\in\partial a\}$. Following the fact that the addition of TNS and TNOs is accomplished by a direct sum~\cite{Schollwoeck:2011:DMRG}, we may stack messages $m_{a\rightarrow b}^{(\pm)}$ and the tensors $W_a^{(\pm)}$ from the respective TNOs block-diagonally to obtain messages and tensors on the complete Hamiltonian:\footnote{In writing $W_a$ as an operator-valued matrix, Eq.~\eqref{BPDMRG:alg:nondef:eq:directsum:optensor} shows the example scenario of site $a$ having two neighbors; the general case proceeds analogously.}
\begin{align}
    m_{a\rightarrow b}^{(+)\oplus (-)} &= \left(\begin{array}{c}
         m_{a\rightarrow b}^{(+)} \\ m_{a\rightarrow b}^{(-)}
    \end{array}\right),
    \quad\text{and}
    \label{BPDMRG:alg:nondef:eq:directsum:message}
    \\
    W_a^{(+)\oplus(-)} &= \left(\begin{array}{cc}
        W_a^{(+)} &  \\ & W_a^{(-)}
    \end{array}\right).
    \label{BPDMRG:alg:nondef:eq:directsum:optensor}
\end{align}
The local Hamiltonian of $H^{(+)} \oplus H^{(-)}$ simplifies to the sum of the PSD- and NSD-parts:
\begin{equation}
    \begin{aligned}
        H_a &\equiv H_a^{(+)\oplus (-)} = \cntr\left(W_a^{(+)\oplus (-)}\prod_{b\in\partial a}m_{b\rightarrow a}^{(+)\oplus (-)}\right) \\
        &= H_a^{(+)} + H_a^{(-)}.
    \end{aligned}
    \label{BPDMRG:alg:nondef:eq:local_Hamiltonian:directsum}
\end{equation}
Putting everything together, when optimizing a single site tensor $T_a$, BP-DMRG prescribes calculation of $H_a=H_a^{(+)} + H_a^{(-)}$ and $N_a$ from BP fixed point messages, thus defining the generalized eigenvalue problem
\begin{equation}
    \left(H_a^{(+)} + H_a^{(-)}\right)\ket{T_a} = \lambda N_a\ket{T_a}.
    \label{BPDMRG:alg:nondef:eq:geneig}
\end{equation}
As in normal DMRG, the solution $T_a^\text{(next)}$ is inserted into the TNS, and the algorithm moves on to the next site.

\subsubsection{The condition number of \texorpdfstring{$N_a$}{local environments}}
\label{sec:BPDMRG:alg:cond}

As we alluded to in Sec.~\ref{sec:backgr:DMRG}, the generalized eigenvalue problem of Eq.~\eqref{backgr:DMRG:eq:gen_eigval} may be highly numerically unstable when $N_a$ is ill-conditioned~\cite{Schollwoeck:2011:DMRG}. BP-DMRG faces this problem, too, when employing Eq.~\eqref{BPDMRG:alg:nondef:eq:geneig}. This section will investigate this issue and introduce a way to mitigate it.

\vspace{\baselineskip}

To illustrate ill-conditioning in BP-DMRG, we execute BP-DMRG with the TFI (Eq.~\eqref{BPDMRG:num:eq:TFI}) ($J=1$ and $g\in[3,4]$) on a hexagonal lattice with $2\times 2$ cells. The Ansatz state is a TNS with randomly generated, complex-valued tensors and $\chi=3$ on all virtual edges. We execute three sweeps, where each sweep consists of visiting each node once and solving Eq.~\eqref{BPDMRG:alg:nondef:eq:geneig} to obtain a new site tensor. Starting from 50 equidistant values $g\in[3,4]$, we construct the corresponding Hamiltonians and execute BP-DMRG on each one was three times.

We find that the condition numbers $\kappa(m_{a\rightarrow b})$ of BP fixed point messages grow very large during execution of BP-DMRG.\footnote{In this work, we define the condition number of a matrix $A$ as $\kappa(A) = \norm{A}_2 \cdot \norm{A^{-1}}_2$~\cite{Strang:1980:LinAlg}.} In every sweep, there are messages with $\kappa(m_{a\rightarrow b})\approx 10^{19}$, with $\kappa$ increasing noticeably between the first and the second sweep. Since the local environments $N_a$ are direct sums of the messages, they are similarly ill-conditioned. This renders the BP-DMRG algorithm, as introduced thus far, unusable.

\vspace{\baselineskip}

\begin{figure}
    \begin{center}
        \begin{tikzpicture}
            \input{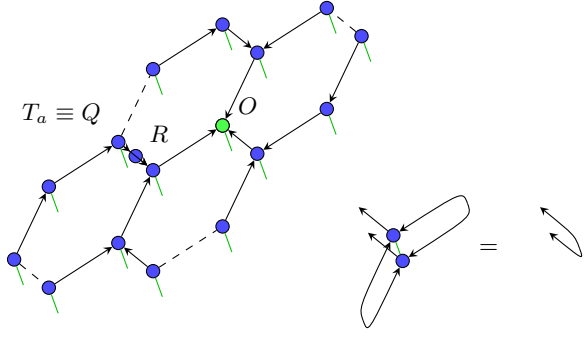}
        \end{tikzpicture}
    \end{center}
    \caption{During tree-gauging, a rooted spanning tree is constructed, and the TNS is orthogonalized along it using QR-decompositions. The node $a$ is decomposed into $T_a=QR$, where $Q$ becomes the new site tensor and $R$ is absorbed in the tensor that lies closer to the root, with the root labeled $O$ in green. Non-tree edges are ignored, as indicated by dashed lines. Tree gauging is able to move $N_a$ closer to the identity, thereby providing improved conditioning of the generalized eigenvalue problem of BP-DMRG (Eq.~\eqref{BPDMRG:alg:nondef:eq:geneig}). Orthogonalization remains approximate, however.}
    \label{BPDMRG:alg:cond:fig:gauging_dist_to_id}
\end{figure}

Just as in the one-dimensional case, the conditioning of the generalized eigenvalue problem may be improved by using appropriate gauging of the TNS. Although there is no easily constructed site-canonical form on arbitrary TNS, one can construct approximate canonical forms. One such form is the ``tree gauge,'' introduced in Ref.~\cite{Gray:2024:HyperApproxContraction}. It is constructed by forming a rooted spanning tree of the TNS graph and decomposing site tensors using QR decomposition. Starting at the leaves of the spanning tree, the isometries $Q$ become the new site tensors, while the matrices $R$ are passed upstream, until the process terminates at the root $O$. Edges of the TNS that are not present in the spanning tree are ignored (see Fig.~\ref{BPDMRG:alg:cond:fig:gauging_dist_to_id}). This procedure is able to improve the conditioning of the environments by moving $N_a$ closer to the identity;\footnote{Closeness of an environment $N_a$ to the identity is quantified by normalizing $N_a$ such that its trace is equal to the identity's. Let $k$ be the size of the vector space that $N_a\in\mathbb{C}^{k\times k}$ acts on, then we take the distance of $N_a$ from the identity to be $d = \norm*{\frac{k}{\tr(N_a)}N_a - I}_F$. Testing local environments from the aforementioned BP-DMRG runs, we find that tree gauging reduces $d$ by one to two orders of magnitude.} tree gauging still represents an approximation in the presence of loops, however (see Fig.~\ref{BPDMRG:alg:cond:fig:gauging_dist_to_id}).

Tree gauging is added to BP-DMRG as a preprocessing step before constructing $H_a$ and $N_a$. The current TNS is gauged with $a$ as the orthogonality center, and the BP iteration is then executed.

\vspace{\baselineskip}

\begin{figure}
    \centering
    \includegraphics[width=\linewidth]{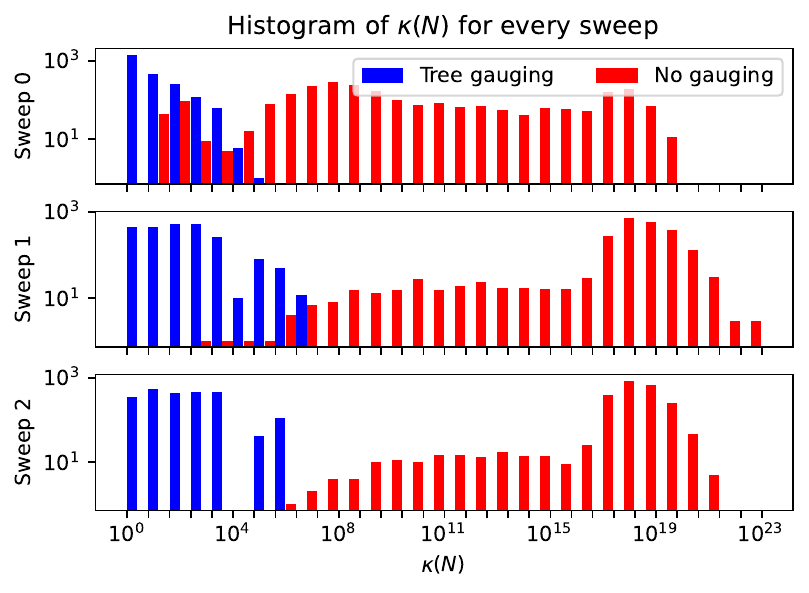}
    \caption{This figure shows histograms of $\kappa(N_a)$ from 300 runs of BP-DMRG with three sweeps each on a $2\times 2$ unit cell hexagonal graph, amounting to 14400 environments $N_a$. Without gauging, $\kappa(N_a)$ grows significantly between the first and the second sweep. With tree-gauging, however, this growth is constrained; the majority of local environments always satisfy $\kappa(N_a)<10^4$. Tikhonov normalization is employed for severely ill-conditioned $N_a$, in effect capping the growth of condition numbers.}
    \label{BPDMRG:alg:cond:fig:environment_cond}
\end{figure}

Additionally, we improve the conditioning of numerically unstable generalized eigenvalue problems using ``Tikhonov regularization'' of $N_a$~\cite{Ghojogh:2023:GenEigvalProblems, Hansen:1998:RankDeficientIllPosedProblems}. It consists of strengthening the main diagonal of $N_a$ according to
\begin{equation}
    N_a \mapsto N_a + \varepsilon I.
    \label{BPDMRG:alg:cond:eq:tikhonov}
\end{equation}
We choose $\varepsilon = 10^{-6}$ and employ Tikhonov-regularization if $\kappa(N_a)>10^6$. The effect of both tree gauging and Tikhonov regularization on environment conditioning is shown in Fig.~\ref{BPDMRG:alg:cond:fig:environment_cond}.

\subsection{Numerical Results}
\label{sec:BPDMRG:num}

Let us recapitulate the BP-DMRG algorithm as introduced thus far. Assume a Hamiltonian $H$ is at hand, which is split up in its positive- and negative-semidefinite parts $H^{(+)}$ and $H^{(-)}$, both of which are given in TNO format. A preset number of sweeps through the TNS Ansatz state is performed, with each sweep involving a single variational optimization of the site tensors in the Ansatz. Optimizing one TNS tensor $T_a$ consists of the following steps:
\begin{itemize}
    \item Apply tree-gauging to $\ket{\psi}$, with the orthogonality center in $a$.
    \item Find the BP fixed points on $\braket{\psi|H^{(+)}|\psi}$, $\braket{\psi|H^{(-)}|\psi}$, and $\braket{\psi|\psi}$.
    \item Define local Hamiltonians $H_a^{(\pm)}$ and the local environment $N_a$ from BP fixed-point messages.
    \item Solve the generalized eigenvalue problem in Eq.~\eqref{BPDMRG:alg:nondef:eq:geneig} for the smallest eigenvalue $\lambda$, obtain the new TNS site tensor $T_a^{\text{(next)}}$, and insert it into the Ansatz state. Move on to the next site.
\end{itemize}

In this section, we will discuss the BP-DMRG algorithm in practice. To this end, we discuss simulation results for BP-DMRG on the transverse-field Ising (TFI) model
\begin{equation}
    H = J\sum_{\braket{a,b}}Z_aZ_b + g\sum_aX_a,
    \label{BPDMRG:num:eq:TFI}
\end{equation}
with varying transverse fields, varying Ansatz bond dimensions $\chi\in\{3,6,9\}$, and on different topologies.\footnote{We have tested BP-DMRG on the Heisenberg model, too, and found that it is able to produce ground states with fidelities similar to the ones that we reach on the TFI.} The Ansatz state is initialized with random, complex-valued site tensors $T_a$. Unless otherwise mentioned, three sweeps are executed, with nodes being visited in arbitrary order.

We will discuss the convergence process and show the infidelity of the ground states obtained by BP-DMRG, as well as the relative accuracy of the corresponding ground-state energies.

\subsubsection{Convergence}
\label{sec:BPDMRG:num:conv}

First, we will demonstrate that BP-DMRG converges, and whether this is detectable within the accuracy of BP.

\vspace{\baselineskip}

Every sweep of BP-DMRG yields a new state $\ket{\psi^{(t)}}$. Accordingly, for every sweep, there is an estimate
\begin{equation}
    E^{(t)} = \frac{\braket{\psi^{(t)}|H|\psi^{(t)}}}{\braket{\psi^{(t)}|\psi^{(t)}}}
    \label{BPDMRG:num:conv:eq:energy_estimate}
\end{equation}
of the ground state energy. We choose the residuals $\varepsilon = \abs*{E^{(t+1)}-E^{(t)}}$ as a measure of convergence of BP-DMRG. The energy estimate may be evaluated in two ways.

First, one may use BP contraction and set
\begin{equation}
    E_\text{BP}^{(t)} = \frac{\braket{\psi^{(t)}|H^{(+)}|\psi^{(t)}}_\text{BP} + \braket{\psi^{(t)}|H^{(-)}|\psi^{(t)}}_\text{BP}}{\braket{\psi^{(t)}|\psi^{(t)}}_\text{BP}}. \label{BPDMRG:num:conv:fig:energy_estimate:BP}
\end{equation}
This is our method of choice during BP-DMRG calculations, since BP contraction is more computationally lightweight than exact contraction. We will denote such estimates as $E_\text{BP}^{(t)}$.

Alternatively, energy estimates may be evaluated via exact contraction, yielding
\begin{equation}
    E_\text{exact}^{(t)} = \frac{\cntr(\braket{\psi^{(t)}|H|\psi^{(t)}})}{\cntr(\braket{\psi^{(t)}|\psi^{(t)}})}. \label{BPDMRG:num:conv:fig:energy_estimate:exact}
\end{equation}
This method is more computationally intensive; however, it removes the inaccuracies inherent in BP contraction and benchmarks the behavior of BP-DMRG. We will denote these estimates as $E_\text{exact}^{(t)}$.

\vspace{\baselineskip}

\begin{figure}
    \includegraphics[width=\linewidth]{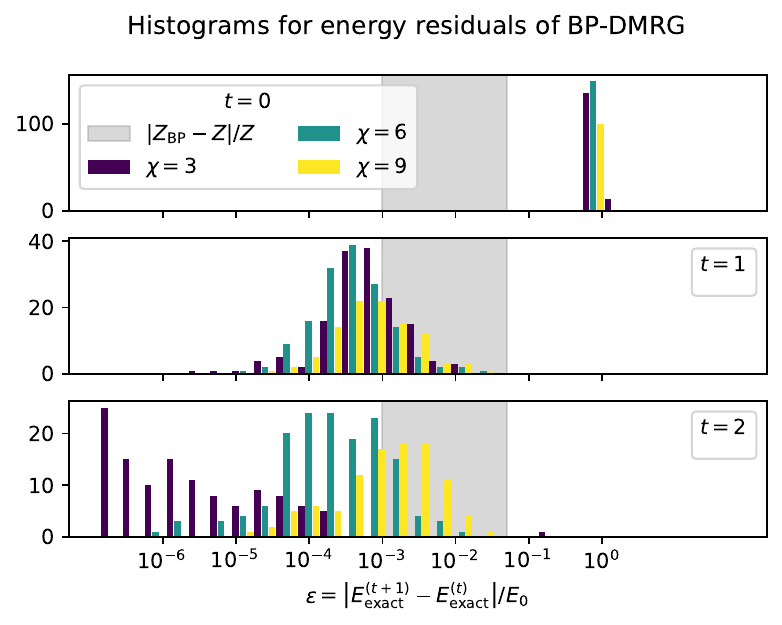}
    \caption{Histograms of relative energy residuals from BP-DMRG for different bond dimensions $\chi$, calculated by exact contraction. Residuals decrease exponentially for $\chi=3$, signaling convergence. For $\chi=9$ (and, to a lesser extent, $\chi=6$), residuals do not decrease much after the initial sweep. The shaded region relates the energy residual to the BP contraction relative accuracy (Fig.~\ref{app:practice:cntr:fig:loop_cntr_approx}). The residuals lie below the contraction error, indicating that convergence within BP accuracy is achieved.}
    \label{BPDMRG:num:conv:fig:E0_exactcntr_everyiter}
\end{figure}

Decreasing residuals $\varepsilon$ signal convergence of BP-DMRG. We evaluate $\varepsilon$ first using exact contraction on hexagonal graphs with $2\times 2$ unit cells. Results for $\chi\in\{3,6,9\}$ are shown in Fig.~\ref{BPDMRG:num:conv:fig:E0_exactcntr_everyiter}. All residuals are initially large, since the initial estimate $E^{(0)}$ is random. It is apparent that $\varepsilon$ decreases exponentially for $\chi=3$, signaling convergence of BP-DMRG.

For $\chi=9$ (and, to a lesser extent, $\chi=6$), residuals do not decrease after the first sweep. Since $\varepsilon$ was calculated using exact contraction, this points to less stable convergence of BP-DMRG for larger bond dimensions.\footnote{This does not mean that BP-DMRG takes longer to converge for larger bond dimensions. We have tested BP-DMRG with 10 sweeps and a bond dimension $\chi=6$, and found that convergence is not simply slower on larger bond dimensions. Instead, energy residuals do not shrink, suggesting that more fundamental obstacles towards convergence are present.}

The same picture emerges when we treat the TFI with the same parameters on a heavy-hexagonal graph with $2\times 2$ unit cells.

\vspace{\baselineskip}

In applications, energy estimates and other observables should be extracted using BP, since it enables more efficient contraction. This raises the question of how the aforementioned convergence stability of BP-DMRG interacts with the contraction accuracy that BP reaches. Phrasing the question more precisely, we are asking if the fluctuations in $E^{(t)}$ lie below the accuracy that BP contraction provides.

To answer this question, we compare the relative accuracy $\abs{Z_\text{BP}-Z}/Z$ of BP contraction to the ratio $\varepsilon/E_0$ of residuals to the true ground state energy (see Fig.~\ref{BPDMRG:num:conv:fig:E0_exactcntr_everyiter}). BP relative contraction accuracies are taken from the results we present in Sec.~\ref{sec:app:practice:cntr}. For $\chi=3$, the relative residuals lie below the BP error. When BP is used to extract quantities of interest, fluctuations caused by BP-DMRG are thus superseded by the inaccuracies of BP contraction; BP-DMRG convergence is more stable than BP contraction can resolve.

\subsubsection{Ground State Energy and Fidelity}
\label{sec:BPDMRG:num:res}

\begin{figure*}
    \begin{center}
        \includegraphics[width=\linewidth]{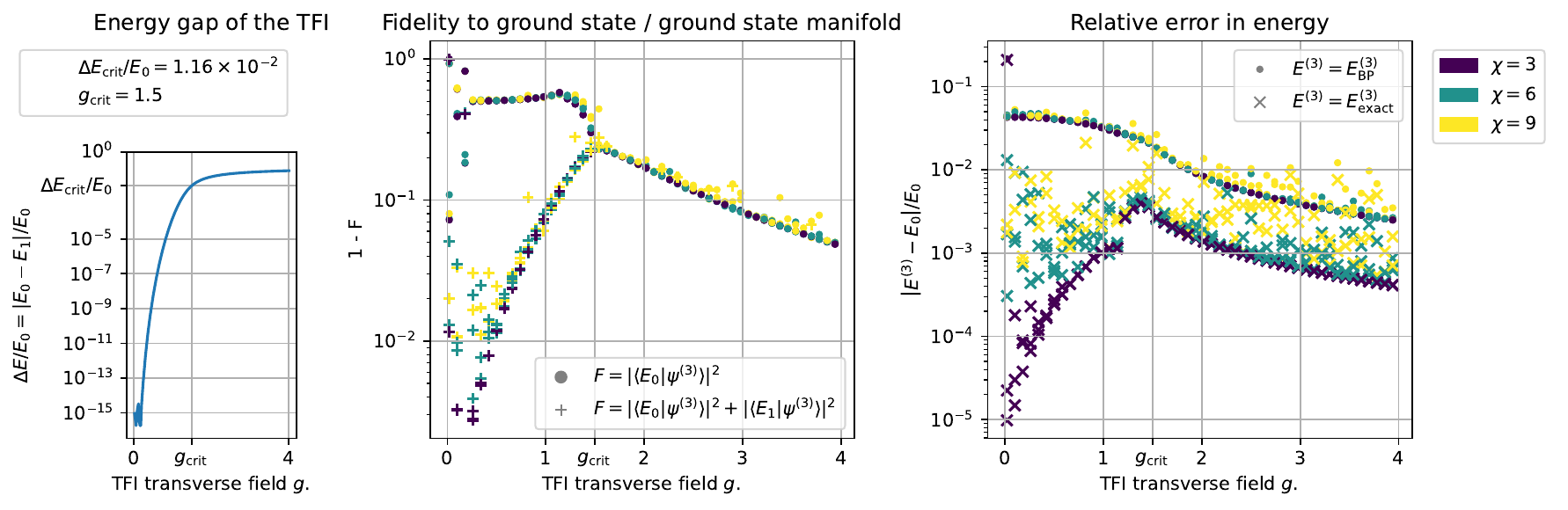}
    \end{center}
    \caption{Energy gap, fidelity to the ground state or the ground state manifold, and energy estimate $E^{(3)}$ after three sweeps, for varying TFI transverse fields. \textbf{(left)} The energy gap between $E_0$ and $E_1$ vanishes for $g\rightarrow 0$, making it harder for BP-DMRG to discriminate between them. \textbf{(middle)} The infidelity of BP-DMRG to the ground state and the ground state manifold, for varying transverse fields. Below $g\approx 1.5$, BP-DMRG cannot discriminate between $\ket{E_0}$ and $\ket{E_1}$. \textbf{(right)} Due to non-discrimination, the relative error in energy increases especially around $g_\text{crit}\approx 1.5$. An additional source of error for $g\le 1.5$ plagues the estimates $E_\text{BP}^{(3)}$, which we will discuss in Sec.~\ref{sec:physBP}. Exact contraction can recover more accurate energy estimates, in line with the fact that the fidelity to the ground state manifold increases for small $g$.}
    \label{BPDMRG:num:res:fig:E0_fidelity}
\end{figure*}

To benchmark BP-DMRG's variational power, we calculate approximate ground states of the TFI for $0\leq g\leq 4$ and $J=1$ on a hexagonal lattice with $2\times 2$ unit cells (Fig.~\ref{BPDMRG:alg:cond:fig:gauging_dist_to_id}). We chose this system since its modest size allows exact diagonalization, while still containing loops of different lengths. Ansatz states are, again, initialized with bond dimensions $\chi\in\{3,6,9\}$ and complex-valued tensors, and three sweeps are executed. The resulting fidelities to the true ground state and energy estimates are shown in Fig.~\ref{BPDMRG:num:res:fig:E0_fidelity}.

\vspace{\baselineskip}

We evaluate the fidelities
\begin{equation}
    \begin{gathered}
        F_0 = \abs{\braket{E_0|\psi^{(3)}}}^2 \quad \text{and}\\ F_{01} = \abs{\braket{E_0|\psi^{(3)}}}^2 + \abs{\braket{E_1|\psi^{(3)}}}^2
    \end{gathered}
    \label{BPDMRG:num:res:eq:fidelities}
\end{equation}
with exact contraction. Throughout this chapter, we will often refer to the ``infidelity'' $1-F$ as a measure of closeness, such that lower values imply closeness to the ground state $\ket{E_0}$ or to $\text{span}\{\ket{E_0}, \ket{E_1}\}$. $\ket{E_0}$ and $\ket{E_1}$ are the ground state and the first excited state, respectively, which are calculated by exact diagonalization along with the energies $E_0$ and $E_1$.

\vspace{\baselineskip}

Let us now relate BP-DMRG results to the properties of TFI ground states. (On the lattice we consider, the TFI ground state is doubly degenerate for $g\lesssim 0.25$. The energy gap $\Delta E = \abs{E_0-E_1}$ increases beyond that for growing transverse fields; see Fig.~\ref{BPDMRG:num:res:fig:E0_fidelity}).

Looking at Fig.~\ref{BPDMRG:num:res:fig:E0_fidelity}(middle), it becomes apparent that all bond dimensions give similar results. More importantly, the results show that BP-DMRG cannot discriminate between $\ket{E_0}$ and $\ket{E_1}$ for small transverse fields. For $g\approx 1.5$, the energy gap is $\Delta E / E_0 \approx 1.16 \times 10^{-2}$, which is approximately the relative accuracy of BP contraction. At this point, the BP-DMRG algorithm delivers a state with $1-F_0\approx 1-F_{01}\approx 0.11$. For smaller $g$, and thus smaller $\Delta E$, BP-DMRG ceases to discriminate between $E_0$ and $E_1$, and the fidelities separate. In this regime, BP-DMRG is able to deliver a state within the manifold $\text{span}\{\ket{E_0}, \ket{E_1}\}$ with improving fidelity $F_{01}$, but $1-F_0\approx 0.5$ consistently, indicating that BP-DMRG converges to superpositions within $\text{span}\{\ket{E_0},\ket{E_1}\}$.

Below $g\lesssim 0.25$, the energy gap vanishes, and the ground state is degenerate. BP-DMRG ceases to yield reliable states, in either measure, for any bond dimension.

Above $g>1.5$, BP-DMRG yields states with growing fidelity to the true ground state. As the energy gap has grown sufficiently, BP-DMRG is able to discriminate between $\ket{E_0}$ and $\ket{E_1}$, such that $F_0\approx F_{01}$. We also tested larger transverse fields $g>4$, i.e., larger energy gaps, and found that the fidelity of BP-DMRG states improves until inaccuracies in the BP iteration within BP-DMRG itself prevent further improvement.

\vspace{\baselineskip}

The deliberations concerning the fidelity of the states $\ket{\psi^{(3)}}$ inform our discussion of the ground state energy estimates $E^{(3)}$ (Fig.~\ref{BPDMRG:num:res:fig:E0_fidelity}(right)). At $g\approx 1.5$, BP-DMRG incurs the largest relative errors of the energy estimate. The explanation for this lies in the interplay between its inability to discriminate within $\text{span}\{\ket{E_0}, \ket{E_1}\}$, and the energy gap $\Delta E$. The point $g\approx 1.5$ is the point with the largest energy gap that BP-DMRG fails to resolve, so it naturally yields the largest relative error of the energy estimate.

For larger transverse fields, the relative energy error decreases again. As the energy gap $\Delta E$ grows, BP-DMRG resolves between $\ket{E_0}$ and $\ket{E_1}$, yielding ground states with higher fidelity and more accurate estimates of $E_0$.

For smaller transverse fields ($g<1.5$), where $\ket{\psi^{(3)}}\in\text{span}\{\ket{E_0},\ket{E_1}\}$, the relative error $|E_\text{exact}^{(3)}-E_0|/E_0$ in energy decreases. This is an effect of the shrinking energy gap: It does not matter whether BP-DMRG is able to discern between $\ket{E_0}$ and $\ket{E_1}$, if the energy gap between the two shrinks; either one will give a good estimate of the ground state energy $E_0$.

Furthermore, exact contraction of $E^{(3)}$ can resolve the convergence instability of BP-DMRG for larger $\chi$ (compare Fig.~\ref{BPDMRG:num:conv:fig:E0_exactcntr_everyiter}). As discussed above, the residuals $\varepsilon$ decrease more slowly for larger bond dimensions during BP-DMRG. This resurfaces in Fig.~\ref{BPDMRG:num:res:fig:E0_fidelity} as fluctuations in $E^{(3)}$ for $\chi\in\{6,9\}$, while BP-DMRG is fully converged after three sweeps for $\chi=3$. We leave the effects of larger bond dimensions on BP-DMRG open for future research.

\vspace{\baselineskip}

\begin{figure*}
    \begin{minipage}{.66\linewidth}
        \includegraphics[width=\linewidth]{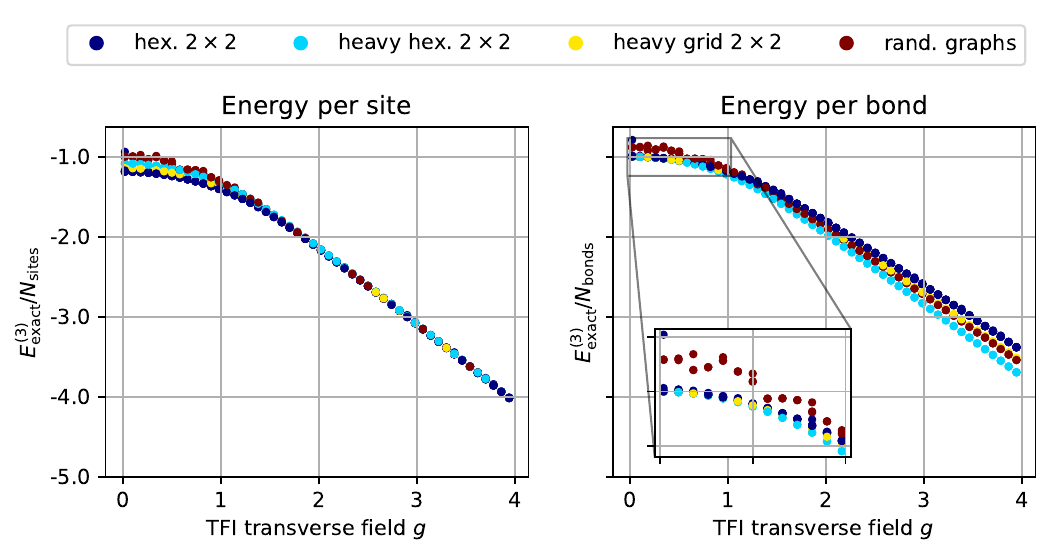}
    \end{minipage}
    \begin{minipage}{.33\linewidth}
        \includegraphics[width=\linewidth]{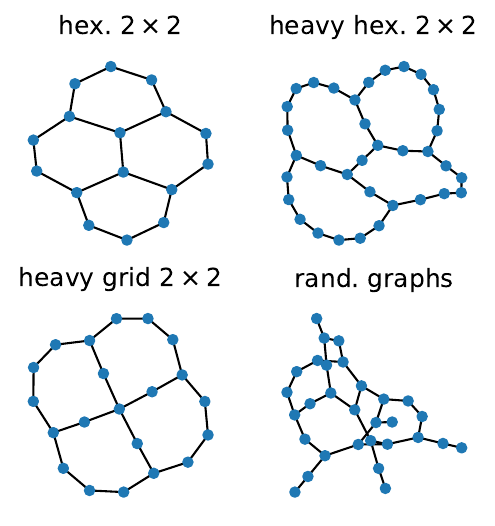}
    \end{minipage}
    \caption{Energy estimates from BP-DMRG per site, and per bond, for different graphs. All Ansatz states have a virtual bond dimension of $\chi=3$. The results for the hexagonal, heavy-hexagonal, and heavy-rectangular systems are the same in the limit of negligible or large transverse field, indicating correct convergence of BP-DMRG. Random graphs diverge slightly for $g\rightarrow 0$. For large transverse fields, BP-DMRG also delivers reliable energy estimates on random graphs.}
    \label{BPDMRG:num:res:fig:energy_density}
\end{figure*}

To substantiate our claim that BP-DMRG can produce ground states on larger, less-structured lattices, we apply BP-DMRG to systems that are too large for exact diagonalization. The accuracy of the results can be assessed by examining the energy per site or per bond. In the limit of large transverse fields, the TFI ground state approaches $\ket{-}^{\otimes N_\text{sites}}$, such that the energy per site is independent of the lattice. In ground states for vanishing transverse fields, all spins are aligned, such that the energy per bond is independent of the lattice. Fig.~\ref{BPDMRG:num:res:fig:energy_density} shows that BP-DMRG can recover this behavior.\footnote{The random graphs examined in Fig.~\ref{BPDMRG:num:res:fig:energy_density} are generated to have a lower-bounded loop length, and nodes with at most three neighbors. They have been generated with the same method that we used in Sec.~\ref{sec:app:practice:cntr} (Fig.~\ref{app:practice:cntr:fig:loop_cntr_approx}), except that the maximum number of neighbors was enforced.}

\section{Belief Propagation on Ground-State TNS and TFI TNO}
\label{sec:physBP}

Having demonstrated the viability of BP-DMRG, we now use our numerical results as a lens to investigate properties of BP itself.

\vspace{\baselineskip}

We established in the previous section that BP-DMRG can find ground states of the TFI. We have also emphasized that, in the spirit of providing a lattice-agnostic solution to the ground state problem, BP should be used ``every step of the way''. That includes the final evaluation of $E^{(t)}=\braket{\psi^{(t)}|H|\psi^{(t)}}$; results were shown in Fig.~\ref{BPDMRG:num:res:fig:E0_fidelity}. BP contraction comes with its own inaccuracies, and we explain in Sec.~\ref{sec:backgr:BP:TN} that these depend on the lattice and the amount of correlation within the TN to be contracted. We will now use the BP-DMRG results for different transverse fields from the previous section -- where the lattice remains unchanged -- to investigate more closely the nature of correlations within $\braket{\psi|H|\psi}$ and how they affect BP contraction accuracy.

The entry point into this discussion is an aspect of Fig.~\ref{BPDMRG:num:res:fig:E0_fidelity}(right) that we have not elucidated yet. Therein, we present two methods for evaluating the ground-state energy estimate $E^{(3)}$, namely $E^{(3)}_\text{BP}$ and $E^{(3)}_\text{exact}$ -- the BP and exact contractions. Since these results have all been obtained on a $2\times 2$ hexagonal lattice, relating one to the other allows us to study BP contraction in relation to the properties of (approximate) ground states of the TFI.

Looking at Fig.~\ref{BPDMRG:num:res:fig:E0_fidelity}(right), we see that $E^{(3)}_\text{BP}$ is consistently less accurate than $E^{(3)}_\text{exact}$ by one order of magnitude for $g>1.5$, indicating a constant relative error between the two. However, for $g<1.5$, the two methods of evaluating $E^{(3)}$ separate: whereas $E^{(3)}_\text{exact}$ moves closer to the exact ground state energy for smaller $g$, $E^{(3)}_\text{BP}$ shows a larger error. This effect is present for all bond dimensions $\chi$. Thus, the BP contraction accuracy changes depending on the TFI transverse field $g$. We will explore this effect in this section. By proposing that the separation is rooted in the TNOs representing the TFI, we aim to demonstrate another aspect of BP that has to be taken into account when using BP in concert with TNS and TNO.

\subsection{The separation of \texorpdfstring{$E^{(3)}_\text{BP}$}{BP contraction} and \texorpdfstring{$E^{(3)}_\text{exact}$}{exact contraction}}
\label{sec:physBP:errs}

Let us first highlight the aforementioned separation between $E^{(3)}_\text{BP}$ and $E^{(3)}_\text{exact}$ more explicitly.

Fig.~\ref{BPDMRG:num:res:fig:E0_fidelity}(right) shows the relative accuracies
\begin{equation}
    \abs*{E_\text{BP}^{(t)}-E_0} / E_0 \quad\text{and}\quad \abs*{E_\text{exact}^{(t)}-E_0} / E_0
    \label{physBP:errs:eq:E0_rel_err}
\end{equation}
that either BP or exact contraction reach in comparison to the true ground state energy $E_0$. We may, however, also compare BP and exact contraction directly, and evaluate the relative accuracy that BP delivers not with respect to the true ground state energy but with respect to $E_\text{exact}^{(3)}$:\footnote{We also use this measure of the BP contraction accuracy for benchmarking in Sec.~\ref{sec:app:practice:cntr}.}
\begin{equation}
    \eta = \abs*{E_\text{BP}^{(t)}-E_\text{exact}^{(t)}} / E_\text{exact}^{(t)} \equiv \abs*{Z_\text{BP}-Z} / Z.
    \label{physBP:errs:eq:BP_rel_err}
\end{equation}

\begin{figure}
    \centering
    \includegraphics[width=\linewidth]{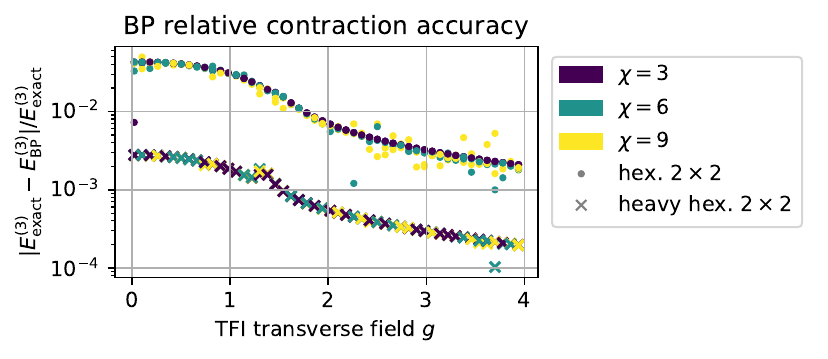}
    \caption{The BP contraction relative accuracy $\eta$ changes with varying transverse fields, although the underlying lattice does not change. While the heavy-hexagonal lattice allows for more accurate contraction across the entire spectrum of $g$, due to the longer loops in a heavy-hexagonal graph, both curves show a growing relative error for a shrinking transverse field $g$.}
    \label{physBP:errs:fig:BP_rel_err}
\end{figure}

The corresponding values for BP-DMRG on the TFI, for hexagonal- or heavy-hexagonal lattices with $2\times 2$ unit cells, are depicted in Fig.~\ref{physBP:errs:fig:BP_rel_err}.

The effect we have claimed before is clearly visible: the BP contraction accuracy of the BP-DMRG estimate depends on the transverse field $g$; namely, it increases as $g$ grows.

In what follows, we will explore possible sources of the separation $\eta$ of $E_\text{BP}^{(3)}$ and $E_\text{exact}^{(3)}$.

Recall that the source of errors is the evaluation of $\braket{\psi^{(3)}|H|\psi^{(3)}}$ using BP. We shall discuss BP contraction as it relates to $\ket{\psi^{(3)}}$ and the TNO $H$, by presenting empirical results for different TN under varying TFI transverse fields $g$.

\subsection{BP contraction of \texorpdfstring{$\braket{E_0|E_0}$ and $\braket{\psi^{(3)}|\psi^{(3)}}$}{TNS norms}}
\label{sec:physBP:TNS}

Two effects overlap within the BP contraction of TNS state norms $\braket{\psi|\psi}$, making it challenging to clearly locate the source of the separation $\eta$.

Firstly, as noted above, Ansatz states with different bond dimensions show different convergence behavior with respect to BP-DMRG, partly hiding the effect we seek to investigate.

Secondly, BP-DMRG aims to transform the Ansatz state into the ground state $\ket{E_0}$ of the TFI Hamiltonian. We have, following Eq.~\eqref{BPDMRG:num:eq:TFI}, $\lim_{g\rightarrow\infty}\ket{E_0}=\ket{-}^{\otimes N}$, with $\ket{-}$ being a product state; on the other end of the parameter range, $\lim_{g\rightarrow 0}\ket{E_0}$ has a high overlap with the manifold $\text{span}\{\ket{0}^{\otimes N},\ket{1}^{\otimes N}\}$. BP contraction is exact on product states,\footnote{This may be proven by going back to Eq.~\eqref{backgr:BP:TN:eq:BP_cntr}. Note first that every message enters exactly two times in $Z_\text{BP}$ (once in the numerator and once in the denominator), and secondly that $\chi=1$ on a product state TNO implies that messages and TNO tensors $T_a$ are scalars. Substituting in Eq.~\eqref{backgr:BP:TN:eq:BP_cntr} reveals that all messages cancel out and $Z_\text{BP}$ simplifies to $Z_\text{BP}=\prod_{a\in V}T_a\equiv Z$. This still holds when the bond dimensions of a product state TNS are enlarged by zero-padding TNS tensors $T_a$ and inserting $I=X^{-1}X$, i.e., by artificially embedding a product state into a larger TNO.} so we expect exact ground states and well-converged BP-DMRG states $\ket{\psi^{(3)}}$ to admit more accurate contraction using BP than randomly generated TNS.

\vspace{\baselineskip}

\begin{figure}
    \centering
    \includegraphics[width=\linewidth]{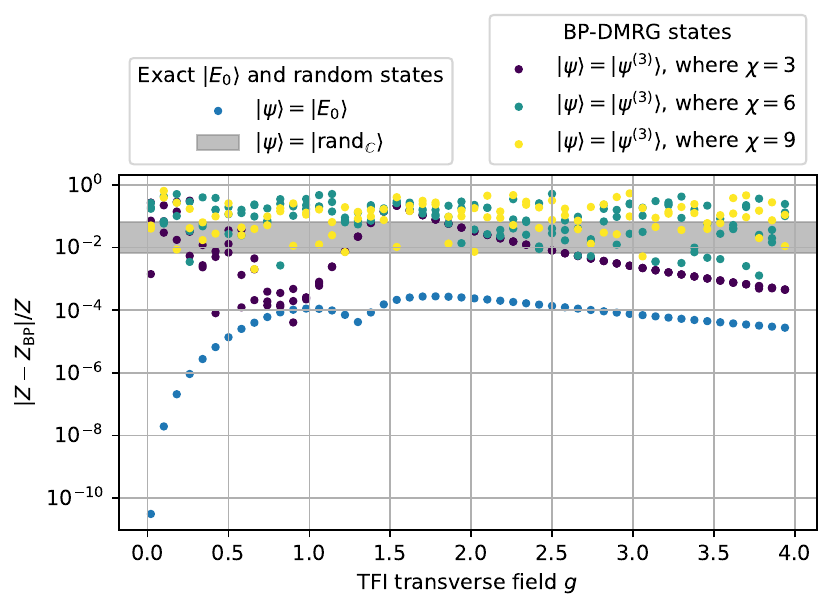}
    \caption{The states $\ket{\psi^{(3)}}$ from BP-DMRG and the exact ground states $\ket{E_0}$ of the TFI admit BP contraction to different degrees of accuracy. The shaded region shows the contraction accuracy on randomly generated TNS on hexagonal graphs (compare with Fig.~\ref{app:practice:cntr:fig:loop_cntr_approx}).}
    \label{physBP:TNS:fig:BP_norm_rel_err}
\end{figure}

These effects notwithstanding, we shall discuss the BP contraction accuracy for the TFI ground state and BP-DMRG state norms. Fig.~\ref{physBP:TNS:fig:BP_norm_rel_err} shows the relative accuracy of BP contraction of $\braket{\psi^{(3)}|\psi^{(3)}}$ and $\braket{E_0|E_0}$, with the exact ground state $\ket{E_0}$ on a hexagonal lattice of $2\times 2$ unit cells being obtained from exact diagonalization.\footnote{Once $\ket{E_0}$ is available as statevector from exact diagonalization of the Hamiltonian matrix, we construct a corresponding TNS through successive SVDs. Since this procedure results in TNS with varying bond dimensions, we zero-pad the TNS tensors $T_a$ and re-gauge virtual edges by inserting $I=X^{-1}X$, enlarging all bond dimensions to a size of $\chi=16$.} The shaded region shows, for comparison, one confidence interval of the BP contraction relative accuracy on $\braket{\text{rand}|\text{rand}}$ with randomly generated TNS $\ket{\text{rand}}$.\footnote{We generate random TNS by randomly filling TNO tensors $T_a\in\mathbb{C}^{\chi^{|\partial a|}\times D}$ with complex numbers, and we set $D=2$ to ensure comparability with the states $\ket{E_0}$ and $\ket{\psi^{3}}$. Tensor entries are drawn from standard normal disributions. In the case of complex-valued tensors, both reak and imaginary part are drawn from standard normal distributions.}

Note first that the contraction accuracy on $\braket{E_0|E_0}$ (shown in blue) does confirm that exact ground states admit more accurate contraction than on $\braket{\text{rand}|\text{rand}}$ (shaded in grey). Moreover, BP contraction accuracy does improve for $g\rightarrow 0$ and $g\rightarrow\infty$, as $\ket{E_0}$ approaches product states.

Furthermore, the convergence behavior that we have discussed in Sec.~\ref{sec:BPDMRG:num:conv} reappears in the relative contraction accuracy on state norms. BP-DMRG Ansatz states with $\chi=3$ are well-converged, and can thus be expected to be closer to ground states; accordingly, Fig.~\ref{physBP:TNS:fig:BP_norm_rel_err} shows that the BP-DMRG result states $\ket{\psi^{(3)}}$ with $\chi=3$ often permit more accurate BP contraction than random states. This is the case particularly for $g\gtrsim 2.5$ and $0.5<g<1$, where we have found BP-DMRG to perform best (see Sec.~\ref{sec:BPDMRG:num:res}). Conversely, BP-DMRG does not converge well for $\chi\in\{6,9\}$, and the respective states $\ket{\psi^{(3)}}$ can only be BP-contracted with the accuracy of random states.

\vspace{\baselineskip}

We do not, however, take the results presented in Fig.~\ref{physBP:TNS:fig:BP_norm_rel_err} to amount to conclusive evidence that the sought-after separation $\eta$ of $E_\text{BP}^{(3)}$ and $E_\text{exact}^{(3)}$ is rooted in the BP contraction accuracy of the states $\ket{\psi^{(3)}}$. The separation is present in equal amounts for all bond dimensions investigated, while Fig.~\ref{physBP:TNS:fig:BP_norm_rel_err} shows different behavior for different bond dimensions. Moreover, $\eta$ is monotonically decreasing for a growing transverse field $g$, while BP contraction accuracies either do not show a trend ($\chi\in\{6,9\}$) or follow BP-DMRG ($\chi=3$), making it hard to see a correlation between $\eta$ and BP contraction accuracy on BP-DMRG states.

\subsection{BP contraction of \texorpdfstring{$\tr(H^{(+)})$}{TNO traces}}
\label{sec:physBP:TNO}

Let us turn our attention to the second component of $\braket{\psi^{(3)}|H|\psi^{(3)}}$: the operator $H$.

\begin{figure}
    \begin{center}
        \begin{tikzpicture}
            \pgfmathsetmacro{\k}{.75}
            \pgfmathsetmacro{\d}{.5*\k}
            \node (T0) at (0, 0) [tponode] {};
\node (T1) at ($(T0)+(-37:1.5*\k)$) [tponode] {};
\node (T2) at ($(T1)+(20:1.5*\k)$) [tponode] {};

\node (label) at ($(T0)+(-0.8*\k, -0.45*\k)$) {$H=$};

\coordinate (T2eastlabel) at (T2.east |- label);
\node (label2) [anchor=west] at ($(T2eastlabel)+(0.5*\k, 0)$) {$\Rightarrow \tr(H)=$};

\coordinate (label2eastT0) at (label2.east |- T0);
\node (T3) at ($(label2eastT0)+(0.5*\k, 0)$) [tponode] {};
\node (T4) at ($(T3)+(-37:1.5*\k)$) [tponode] {};
\node (T5) at ($(T4)+(20:1.5*\k)$) [tponode] {};

\draw (T0) -- (T1);
\draw (T1) -- (T2);
\draw (T3) -- (T4);
\draw (T4) -- (T5);

\draw ($(T0)+(0, \d)$) -- (T0) -- ($(T0)+(0, -\d)$) [physedge];
\draw ($(T1)+(0, \d)$) -- (T1) -- ($(T1)+(0, -\d)$) [physedge];
\draw ($(T2)+(0, \d)$) -- (T2) -- ($(T2)+(0, -\d)$) [physedge];

\draw [physedge, rounded corners=3pt] (T3) -- ($(T3)+(0, \d)$) -- ($(T3)+(0.35*\k, \d)$) -- ($(T3)+(0.35*\k, -\d)$) -- ($(T3)+(0, -\d)$) -- (T3);
\draw [physedge, rounded corners=3pt] (T4) -- ($(T4)+(0, \d)$) -- ($(T4)+(0.35*\k, \d)$) -- ($(T4)+(0.35*\k, -\d)$) -- ($(T4)+(0, -\d)$) -- (T4);
\draw [physedge, rounded corners=3pt] (T5) -- ($(T5)+(0, \d)$) -- ($(T5)+(0.35*\k, \d)$) -- ($(T5)+(0.35*\k, -\d)$) -- ($(T5)+(0, -\d)$) -- (T5);
        \end{tikzpicture}
    \end{center}
    \caption{Every site tensor $W_a$ of a TNO has two physical legs, relating to the domain and the image of the linear map $H$. Connecting the respective legs of every site $a\in V$ gives a TN that evaluates to the trace of $H$. We use such TN to investigate the relation between $H$ and BP contraction accuracy.}
    \label{physBP:TNO:fig:H_trace}
\end{figure}
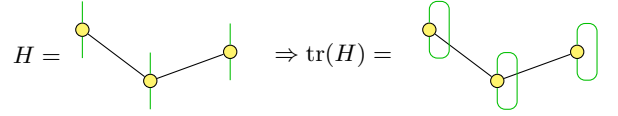

In order to assess the impact that the TFI Hamiltonian has on BP contraction accuracy, we shall look at BP contraction accuracies on TN that represent the trace $\tr(H^{(+)})$ of the Hamiltonian. Specifically, since BP-DMRG necessitates splitting $H$ into its positive- and negative-semidefinite part $H^{(\pm)}$, it suffices to consider solely $H^{(+)}$ in this section. With the TNO $H^{(+)}$ at hand, such a TN is easily obtained by connecting the two physical edges of any TNO tensor $W_a$ (see Fig.~\ref{physBP:TNO:fig:H_trace}).

\begin{figure}
    \centering
    \includegraphics[width=\linewidth]{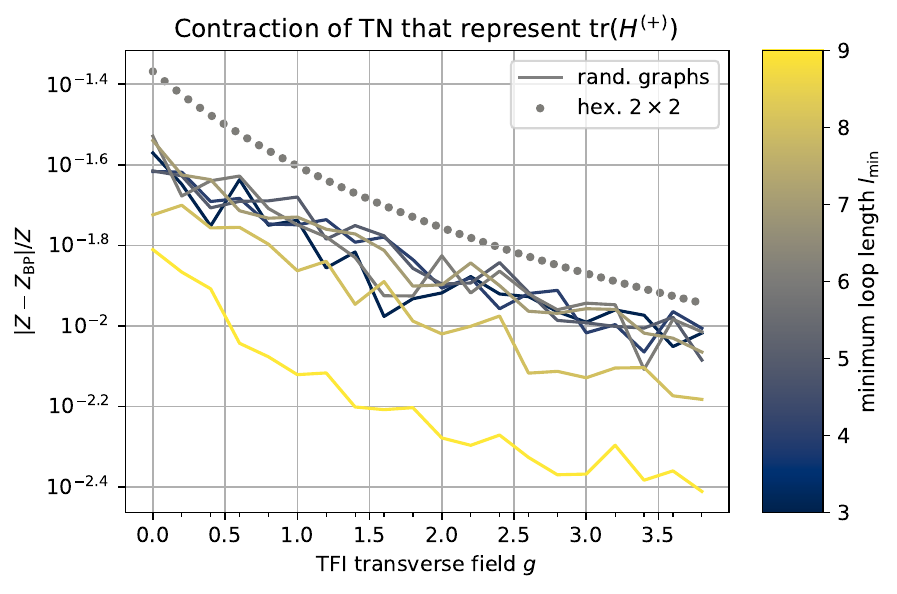}
    \caption{BP contraction accuracy of the trace $\tr(H^{(+)})$ of the positive-semidefinite part of the TFI Hamiltonian, on the hexagonal $2\times 2$ lattice and random lattices. The results for random lattices have been binned in intervals of $0.2$ to suppress noise. While BP contraction accuracy does increase as the loop length within the lattices increases, as we have seen before (Fig.~\ref{backgr:BP:QMBP:fig:hex22_norm}), $\tr(H^{(+)})$ is clearly not well-approximated using BP.}
    \label{physBP:TNO:fig:H_trace_rel_err}
\end{figure}

Fig.~\ref{physBP:TNO:fig:H_trace_rel_err} shows the accuracy with which BP is able to contract $\tr(H^{(+)})$. We look at two different kinds of lattices: the $2 \times 2$ unit cell hexagonal lattice, since it is our BP-DMRG test bed, and the random lattices with minimum loop length from Sec.~\ref{sec:app:practice:cntr} (Fig.~\ref{app:practice:cntr:fig:loop_cntr_approx}).

As we discuss in Sec.~\ref{sec:backgr:BP:TN}, here again lattices with longer loops admit more accurate BP contraction. Additionally, contraction accuracy on both random lattices and the hexagonal $2\times 2$ lattice increases for growing transverse fields.

The growth of contraction accuracy on $\tr(H^{(+)})$ for larger transverse fields mirrors the closing separation $\eta$ (Fig.~\ref{physBP:errs:fig:BP_rel_err}): as $\abs{E_\text{BP}^{(3)} - E_\text{exact}^{(3)}} / E_\text{exact}^{(3)}$ becomes smaller for large $g$, so does $\abs{Z_\text{BP}-Z}/Z$ for the TFI trace $\tr(H^{(+)})$. This underpins our proposition that the TNOs of the transverse field Ising model cause the separation $\eta$ between $E_\text{BP}^{(3)}$ and $E_\text{exact}^{(3)}$, which makes evaluation of BP-DMRG results -- using BP itself -- challenging. We discuss the construction of the TNO for the TFI in more detail in Sec.~\ref{sec:app:MPO}.


\section{Conclusion and Outlook}
\label{sec:close}

In this work, we have introduced BP-DMRG, an algorithm for finding ground-state TNS of many-body systems. In the spirit of providing a lattice-agnostic solution, we introduced BP into DMRG. This brings with it both opportunities and challenges.

BP-DMRG presents an opportunity by admitting an alternative contraction method to DMRG. We have shown that it can enrich ground state search beyond the availability of canonical forms and highly structured lattices. While acknowledging that fidelities between $F\approx 0.9$ and $F\approx 0.99$ may be too low to study the properties of a given system, BP-DMRG may construct priors for other methods, such as imaginary-time evolution.

Nevertheless, applying BP-DMRG to an unknown system is challenging. We have demonstrated, for example, that the performance of BP-DMRG is highly sensitive to both the bond dimension and the energy gap of the TFI. The required bond dimensions, in turn, depend on the entanglement within a state, making it unclear how well BP-DMRG can be applied to systems whose properties are unknown.

This point brings us to what BP-DMRG performance reveals about BP in the context of quantum many-body physics. Having incorporated BP into ground state search, it has clearly proven useful, yet we show in Sec.~\ref{sec:physBP} that the TNOs of the particle-decay construction are not amenable to accurate contraction using BP, as our results evaluating $\tr(H^{(+)})$ indicate.

Our results serve both as a demonstration of the power of BP and of its limitations. BP can be applied to ground-state search, yielding a highly flexible algorithm with respect to the underlying lattice. However, this comes at the cost of reduced accuracy because BP is an uncontrolled approximation in the presence of loops.

\vspace{\baselineskip}

We see many directions for future research. As we have only considered single-site DMRG so far, a natural extension is two-site BP-DMRG, which we expect to work analogously to the algorithm presented here. This could yield an algorithm with greater variational power, since bond dimensions could be adjusted dynamically; we have seen, though, that larger bond dimensions degrade convergence of single-site BP-DMRG. A more sophisticated modification presents itself in the incorporation of Loop Series Expansion into BP-DMRG, which we expect to yield more accurate contraction and thus more granular site tensor updates~\cite{Evenbly:2026:LoopSeriesExpansionPublished}.

In this work, we use the simplest BP iteration schedule. More sophisticated schedules are possible, of course, which could increase BP stability and accelerate convergence~\cite{Elidan:2006:ResidualBeliefPropagation, Koller:2009:ProbabilisticGraphicalModels, Wainwright:2003:TreeReParam}.

Lastly, the relation between BP contraction accuracy on $\braket{\psi|\psi}$, $\tr(H^{(+)})$ and on $\braket{\psi|H^{(+)}|\psi}$ remains poorly understood. Shedding light on why BP delivers accurate or inaccurate results for any of these TNs and how these effects conspire would greatly benefit not just BP-DMRG but other applications of BP in quantum many-body physics, too.

\bibliography{articles,books,preprints,misc}

\appendix

\section{Tensor Network Operators}
\label{sec:app:MPO}

Our BP-DMRG algorithm makes heavy use of tensor network operators (TNOs); we have introduced these as TNs with physical legs, which are the tensor network representations of a Hamiltonian (Eq.~\eqref{backgr:BP:QMBP:eq:TNO}). In this section, we will show how TNOs may be constructed for Hamiltonians that are defined as a sum of operator strings, using the ``particle-decay construction''~\cite{Bridgeman:2017:InterpretativeDance}.

Consider the TFI Hamiltonian again (Eq.~\eqref{BPDMRG:num:eq:TFI})
\begin{equation}
    H = J\sum_{\braket{a,b}}Z_aZ_b + g \sum_a X_a.
    \label{app:MPO:eq:TFI}
\end{equation}
$H$ is defined as the sum of operator strings $J Z_a Z_b$ and $g X_a$, which cover all pairwise interactions and all couplings to the transverse field $g$. Given the lattice on which the system is defined, we now construct the TNO from the images of particles traversing the graph, ``decaying'' along the way. Each path that a particle may take corresponds to an operator string (``decay chain''), and lattice sites traversed contribute operators to the respective operator chain. A particle may decay between lattice sites, contributing non-identity operators; if a particle traverses an edge without decaying, an identity is added to the operator chain.

\begin{figure}
    \begin{center}
        \begin{tikzpicture}
            \input{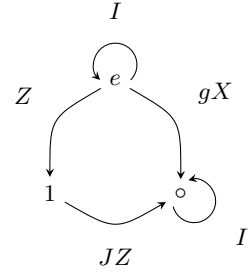}
        \end{tikzpicture}
    \end{center}
    \caption{The particle-decay scheme of the TFI (Eq.~\eqref{app:MPO:eq:TFI}), written as finite-state automaton. There are three particle states: $e$ is the excited state, 1 the decay state, and $\circ$ is the vacuum state. A decay may occur when a particle traverses an edge, denoted by directed edges in the diagram. Operators next to the transition denote the operator that is added to the operator string during that decay. Complete decay chains yield operator strings of the Hamiltonian.}
    \label{app:MPO:fig:finitestate}
\end{figure}

The TFI necessitates decay chains with three particle states: An excited state (``$e$''), an intermediate state (``1''), and a vacuum state (``$\circ$''). The two types of operator strings in the TFI necessitate two different decay chains: The two-step decay $e \rightarrow 1 \rightarrow \circ$ encodes $JZ_aZ_b$, while $e \rightarrow \circ$ encodes $gX_a$. This may be succinctly visualized as a finite-state automaton (see Fig.~\ref{app:MPO:fig:finitestate}).

\begin{figure*}
    \begin{minipage}{.4\textwidth}
    \textbf{(a)}

    \begin{center}
        \begin{tikzpicture}[
            arrow/.style = {->, shorten >=1, shorten <=1}
        ]
            \input{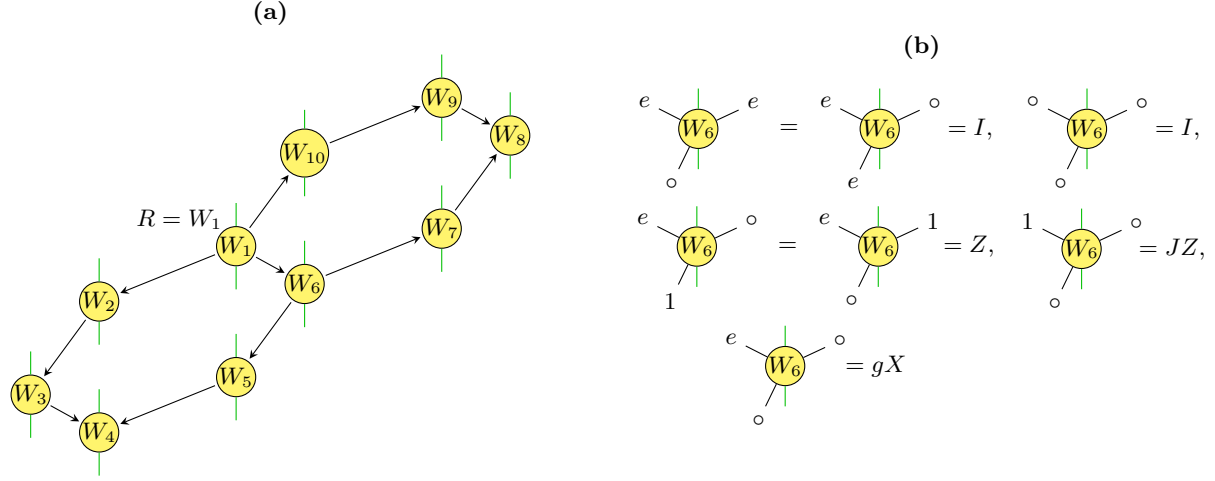}
        \end{tikzpicture}
    \end{center}
    \end{minipage}
    \begin{minipage}{.549\textwidth}
        \textbf{(b)}

        \begin{center}
            \begin{tabular}{cc}
                \begin{tikzpicture}
                    \pgfmathsetmacro{\k}{.8}
                    \pgfmathsetmacro{\d}{\k/1.5}
                    \node (W6) at (0,0) [tponode] {$W_6$};
                    \node (W7) at ($(W6)+(25:\k)$) [] {$e$};
                    \node (W1) at ($(W6)+(153:\k)$) [] {$e$};
                    \node (W5) at ($(W6)+(-116:\k)$) [] {$\circ$};
                    \foreach \i in {7, 1, 5}{\draw (W6) -- (W\i);}
                    \coordinate (leg) at ($(W6)+(0,\d)$);
                    \coordinate (legadj) at ($(W6)+(0,-\d)$);
                    \draw [physedge] (leg) -- (W6) -- (legadj);
                    \node (eq) at ($(W6)+(\k*1.5,0)$) [] {$=$};
                    \node (WW6) at ($(eq)+(\k*1.5,0)$) [tponode] {$W_6$};
                    \node (WW7) at ($(WW6)+(25:\k)$) [] {$\circ$};
                    \node (WW1) at ($(WW6)+(153:\k)$) [] {$e$};
                    \node (WW5) at ($(WW6)+(-116:\k)$) [] {$e$};
                    \foreach \i in {7, 1, 5}{\draw (WW6) -- (WW\i);}
                    \coordinate (leg) at ($(WW6)+(0,\d)$);
                    \coordinate (legadj) at ($(WW6)+(0,-\d)$);
                    \draw [physedge] (leg) -- (WW6) -- (legadj);
                    \node (math) at ($(WW6)+(\k*1.5,0)$) [] {$=I$,};
                \end{tikzpicture}
                &
                \begin{tikzpicture}
                    \pgfmathsetmacro{\k}{.8}
                    \pgfmathsetmacro{\d}{\k/1.5}
                    \node (W6) at (0,0) [tponode] {$W_6$};
                    \node (W7) at ($(W6)+(25:\k)$) [] {$\circ$};
                    \node (W1) at ($(W6)+(153:\k)$) [] {$\circ$};
                    \node (W5) at ($(W6)+(-116:\k)$) [] {$\circ$};
                    \foreach \i in {7, 1, 5}{\draw (W6) -- (W\i);}
                    \coordinate (leg) at ($(W6)+(0,\d)$);
                    \coordinate (legadj) at ($(W6)+(0,-\d)$);
                    \draw [physedge] (leg) -- (W6) -- (legadj);
                    \node (math) at ($(W6)+(\k*1.5,0)$) [] {$=I$,};
                \end{tikzpicture}
                \\
                
                \begin{tikzpicture}
                    \pgfmathsetmacro{\k}{.8}
                    \pgfmathsetmacro{\d}{\k/1.5}
                    \node (W6) at (0,0) [tponode] {$W_6$};
                    \node (W7) at ($(W6)+(25:\k)$) [] {$\circ$};
                    \node (W1) at ($(W6)+(153:\k)$) [] {$e$};
                    \node (W5) at ($(W6)+(-116:\k)$) [] {1};
                    \foreach \i in {7, 1, 5}{\draw (W6) -- (W\i);}
                    \coordinate (leg) at ($(W6)+(0,\d)$);
                    \coordinate (legadj) at ($(W6)+(0,-\d)$);
                    \draw [physedge] (leg) -- (W6) -- (legadj);
                    \node (eq) at ($(W6)+(\k*1.5,0)$) [] {$=$};
                    \node (WW6) at ($(eq)+(\k*1.5,0)$) [tponode] {$W_6$};
                    \node (WW7) at ($(WW6)+(25:\k)$) [] {1};
                    \node (WW1) at ($(WW6)+(153:\k)$) [] {$e$};
                    \node (WW5) at ($(WW6)+(-116:\k)$) [] {$\circ$};
                    \foreach \i in {7, 1, 5}{\draw (WW6) -- (WW\i);}
                    \coordinate (leg2) at ($(WW6)+(0,\d)$);
                    \coordinate (legadj2) at ($(WW6)+(0,-\d)$);
                    \draw [physedge] (leg2) -- (WW6) -- (legadj2);
                    \node (math) at ($(WW6)+(\k*1.5,0)$) [] {$=Z$,};
                \end{tikzpicture}
                &
                \begin{tikzpicture}
                    \pgfmathsetmacro{\k}{.8}
                    \pgfmathsetmacro{\d}{\k/1.5}
                    \node (W6) at (0,0) [tponode] {$W_6$};
                    \node (W7) at ($(W6)+(25:\k)$) [] {$\circ$};
                    \node (W1) at ($(W6)+(153:\k)$) [] {1};
                    \node (W5) at ($(W6)+(-116:\k)$) [] {$\circ$};
                    \foreach \i in {7, 1, 5}{\draw (W6) -- (W\i);}
                    \coordinate (leg) at ($(W6)+(0,\d)$);
                    \coordinate (legadj) at ($(W6)+(0,-\d)$);
                    \draw [physedge] (leg) -- (W6) -- (legadj);
                    \node (math) at ($(W6)+(\k*1.5,0)$) [] {$=JZ$,};
                \end{tikzpicture}
                \\

                \begin{tikzpicture}
                    \pgfmathsetmacro{\k}{.8}
                    \pgfmathsetmacro{\d}{\k/1.5}
                    \node (W6) at (0,0) [tponode] {$W_6$};
                    \node (W7) at ($(W6)+(25:\k)$) [] {$\circ$};
                    \node (W1) at ($(W6)+(153:\k)$) [] {$e$};
                    \node (W5) at ($(W6)+(-116:\k)$) [] {$\circ$};
                    \foreach \i in {7, 1, 5}{\draw (W6) -- (W\i);}
                    \coordinate (leg) at ($(W6)+(0,\d)$);
                    \coordinate (legadj) at ($(W6)+(0,-\d)$);
                    \draw [physedge] (leg) -- (W6) -- (legadj);
                    \node (math) at ($(W6)+(\k*1.5,0)$) [] {$=gX$};
                \end{tikzpicture}
            \end{tabular}
        \end{center}
    \end{minipage}
    \caption{The TNO of the TFI can be constructed in an efficient way using the particle decay construction. \textbf{(a)} Consider the TFI on a hexagonal graph with two unit cells. In this example, we choose node 1 as the root $R$, and thus as the source of excited particles. Arrows denote the downstream direction of the DAG. \textbf{(b)} All non-zero components of the TNO tensor $W_6$; it is most easily defined by conceptualizing it as an operator-valued three-legged tensor.}
    \label{app:MPO:fig:hex21_tfi}
\end{figure*}

The TNO tensors $W_a$ encode the decay chains of particles. Since this work is concerned with arbitrary (possibly loopy) lattices, the tensors $W_a$ are constructed using a directed acyclic graph (DAG) that contains all edges and nodes of the lattice. Starting from a root node in the excited state $e$, particle decay may only occur along the DAG. The example of the TFI on a two-unit-cell hexagonal lattice is depicted in Fig.~\ref{app:MPO:fig:hex21_tfi}.

\section{BP in practice}
\label{sec:app:practice}

\subsection{Contraction of Tensor Networks and \texorpdfstring{$\braket{\psi|\psi}$}{TNS norms}}
\label{sec:app:practice:cntr}

To illustrate BP contraction in practice, we show results for two applications: TN contraction (Eq.~\eqref{backgr:BP:TN:eq:partition_function}), and the evaluation of $\braket{\psi|\psi}$ with $\ket{\psi}$ as TNS.

Both the TN and the TNS are placed on randomly generated lattices. Starting from randomly generated trees with 30 nodes, we add 5 edges to each. Addition of edges is gated by the length of the loops that they create: we impose a minimum loop length $l_{\min}$, and leave the maximum loop length unbounded.

For the case of TN, the graphs are filled with random, positive-valued tensors $T_a\in\mathbb{R}_+^{\chi^{|\partial a|}}$. For the case of TNS $\ket{\psi}$, the tensors $T_a\in\mathbb{C}^{\chi^{\partial a}\times D}$ are filled with random complex numbers, drawn from a standard normal distribution.

We then use Eq.~\eqref{backgr:BP:TN:eq:BP_cntr} to contract these networks. Messages are initialized randomly and entrywise positive. We iterate Eq.~\eqref{backgr:BP:TN:eq:msg_update} (or Eq.~\eqref{backgr:BP:QMBP:eq:msg_update}) for all edges and directions until the BP fixed point is found.\footnote{The halting criterion we adopt in this work is the maximum message residual $\varepsilon=\max_{(a,b)\in E^\leftrightarrow}\varepsilon_{(a,b)}$, where $\varepsilon_{(a,b)}=\norm{m_{a\rightarrow b}^{(t+1)}-m_{a\rightarrow b}^{(t)}}$~\cite{Guo:2023:BlockBP}. In all simulations for this work, we iterate until $\varepsilon \leq 10^{-10}$.\label{app:practice:cntr:fn:halting_criterion}}

\begin{figure*}
    \begin{center}
        \begin{minipage}{.44\textwidth}
            \includegraphics[width=\linewidth]{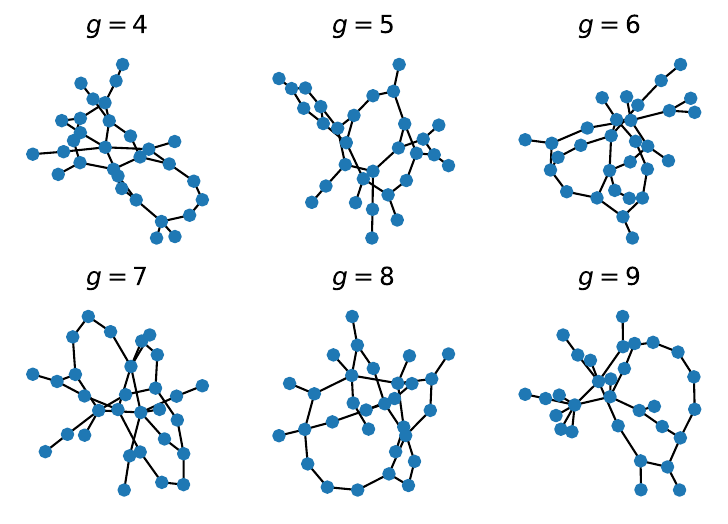}
        \end{minipage}
        \begin{minipage}{.54\textwidth}
            \includegraphics[width=\linewidth]{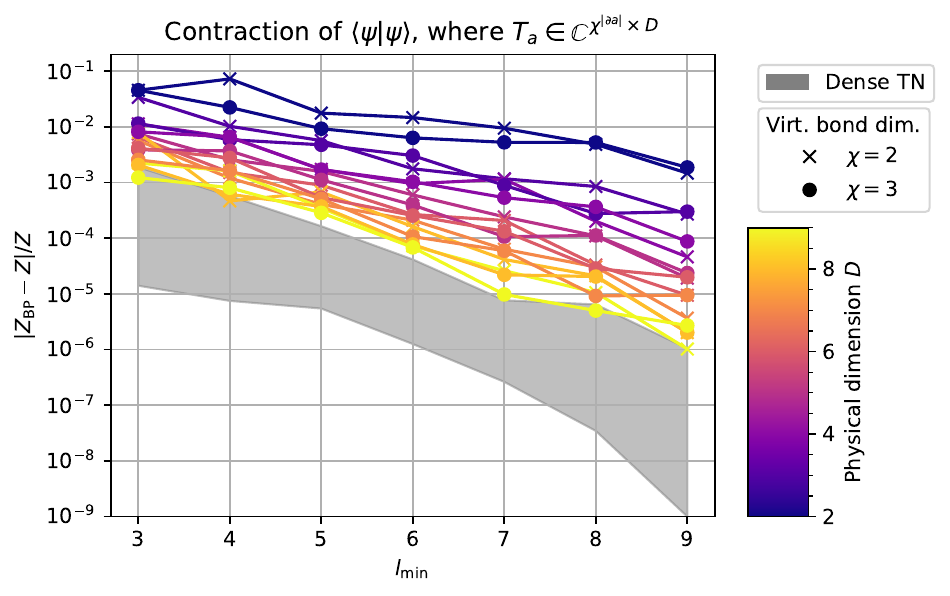}
        \end{minipage}
    \end{center}
    \caption{The BP algorithm is able to contract TNS with varying degrees of accuracy, depending on the structure of the underlying lattice and the size $D$ of the local Hilbert spaces. To show this, we use Eq.~\eqref{backgr:BP:TN:eq:BP_cntr} to contract TN with diverse topologies. \textbf{(Left)} Some example graphs for different minimum loop lengths. \textbf{(Right)} For each minimum loop length, we show mean relative contraction accuracies. The shaded region denotes one confidence interval for the relative contraction accuracy of simple TN (with tensors $T_a\in\mathbb{R}_+^{\chi^{|\partial a|}}$), lines show mean contraction accuracies for $\braket{\psi|\psi}$, where the states $\ket{\psi}$ have been randomly generated. While contraction accuracy is clearly increasing for lattices with longer loops, there are drastic differences depending on the TN under consideration.}
    \label{app:practice:cntr:fig:loop_cntr_approx}
\end{figure*}

Visualizations of the graphs and contraction accuracies can be found in Figure~\ref{app:practice:cntr:fig:loop_cntr_approx}. It is clear that BP can contract the TN in question up to high accuracy, depending on the loops in the graph: long loops admit contraction of TN with a relative error up to $10^{-8}$. On the contrary, the presence of even a few short loops degrades the contraction accuracy to as much as $10^{-2}$. Moreover, since BP is an uncontrolled approximation, contraction accuracy varies by as much as two orders of magnitude.

For squared norms $\braket{\psi|\psi}$ of randomly generated TNS $\ket{\psi}$, the contraction accuracy also improves exponentially with increasing loop length. Moreover, larger physical dimensions $D$ of the local Hilbert spaces admit more accurate contraction. The case most relevant for this work, $D=2$, incurs the largest contraction errors for all the loop lengths investigated. We find no influence of the bond dimension $\chi$ on contraction accuracy.

It should be emphasized that the TN tensors contained only positive numbers. For BP to be a means for accurate, rapid contraction, it is necessary that a TN is \textit{sign-preserving}. This criterion is satisfied by $\braket{\psi|\psi}$, because repeated message updates (Eq.~\eqref{backgr:BP:QMBP:eq:msg_update}) preserve the positive-semidefiniteness of the messages, mirroring the fact that $\braket{\psi|\psi}\geq 0$ for all $\ket{\psi}$. Expectation values $\braket{\psi|O|\psi}$ are sign-preserving if they are either positive-semidefinite or negative-semidefinite.

These limitations notwithstanding, BP can be useful for rapid contraction. If the graph structure is kept in mind, and loops are sufficiently large, TN contraction can be executed with known relative accuracy.\footnote{A notable example is Ref.~\cite{Tindall:2024:EfficientEagleSim}, where the authors have demonstrated how BP can be used to accurately simulate a kicked Ising model on a heavy-hexagonal lattice, surpassing the accuracy from experiments on quantum hardware.}

\subsection{Message Initialization}
\label{sec:app:practice:msg}

In Sec.~\ref{sec:backgr:BP:QMBP}, we claimed that BP converges irrespective of whether the initial messages are positive-semidefinite. In order to test whether different methods do indeed converge, and to what contraction accuracy, we test seven different message initialization methods on $\braket{\psi|\psi}$:
\begin{itemize}
    \item \textit{psd}, and \textit{nsd}: Positive- or negative-semidefinite matrices: $m=U\Lambda U^\dagger$, with a random unitary matrix $U$ and eigenvalues $\Lambda$ that are drawn from a uniform distribution over $[0,1]$ or $[-1,0]$, respectively.
    \item \textit{herm}, and \textit{antiherm}: Hermitian- or anti-hermitian matrices: $m=U\Lambda U^\dagger$, with a random unitary matrix $U$ and eigenvalues $\Lambda=\text{diag}(\{\lambda\})$ or $\Lambda=\text{diag}(\{i\lambda\})$, respectively. The eigenvalues $\lambda$ are drawn from a uniform distribution over $[-1,1]$.
    \item \textit{uniform}: Constant-value messages: $m=\frac{1}{\chi^2}1_{\chi\times\chi}$, where $1_{\chi\times\chi}$ is a $\chi\times\chi$-matrix containing only ones. $\chi$ is the bond dimension.
    \item \textit{zerosum}: Messages whose trace evaluates to zero: $m=U\left(\Lambda - \frac{1}{\chi} \sum\Lambda\right)U^\dagger$, where the eigenvalues $\Lambda$ are drawn from a uniform distribution over $[0,1]$ and $\chi$ is the bond dimension.
    \item \textit{randn}: Messages $m\in\mathbb{C}^{\chi\times\chi}$, whose entries are drawn from a normal distribution $\mathcal{N}(0,1)$ in the complex plane.
\end{itemize}

\begin{figure*}
    \begin{center}
        \includegraphics[width=\textwidth]{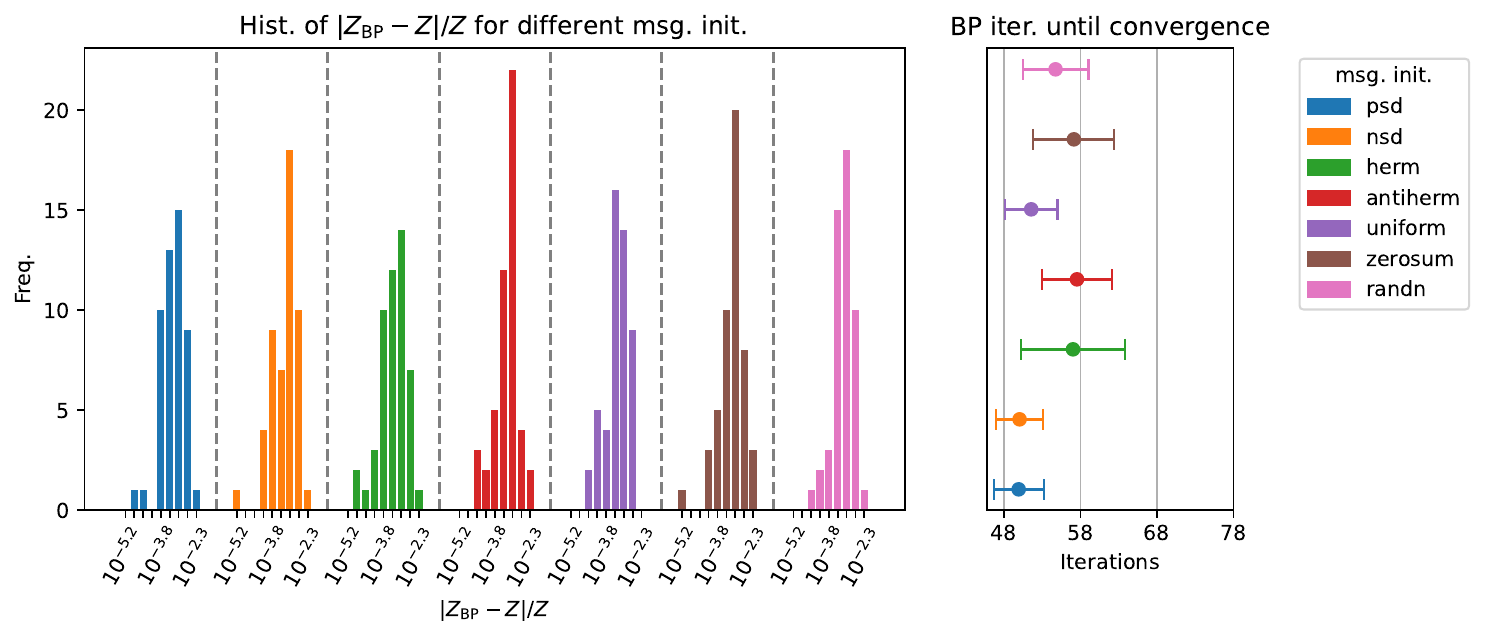}
    \end{center}
    \caption{BP iterations for $\braket{\psi|\psi}$, where $\ket{\psi}$ is a TNS on a heavy-hexagonal graph with $2\times 2$ unit cells. For each message initialization, 50 BP iterations were carried out. All iterations converged to BP fixed points consisting of PSD-messages. The left figure shows histograms of the relative contraction error (smaller values indicate better performance). All message initializations yield fixed points of similar accuracy. On average, for all initializations, $\abs{Z_\text{BP}-Z}/Z\approx 10^{-3.5}$. The figure on the right shows how many iterations BP takes to converge. Although all initializations converge, the methods \lstinline{psd} and \lstinline{nsd} (and, to a lesser extent, \lstinline{uniform}) converge noticeably faster than the others, taking about 10 iterations fewer on average.}
    \label{app:practice:msg:fig:non_psd_messages}
\end{figure*}

We discuss contraction results for all message initialization methods in Fig.~\ref{app:practice:msg:fig:non_psd_messages}. It becomes clear that all initialization methods perform equally well in terms of contraction accuracy. More interestingly, the methods \lstinline{psd} and \lstinline{nsd} converge noticeably faster than the others, taking on average 10 iterations less than \lstinline{herm} and \lstinline{antiherm}. We propose that this is because the \lstinline{psd} and \lstinline{nsd} initialization schemes possess the sign structure of the TN $\braket{\psi|\psi}$. $\braket{\psi|\psi}$ being a stack of $\ket{\psi}$ and its conjugate $\bra{\psi}$, contractions of parts of the TN are PSD, echoing the fact that the messages are PSD matrices. This hints at the fact that $\braket{\psi|\psi}$ is sign-preserving.

Interestingly, the BP iteration also converges to PSD messages when non-PSD initial messages are used, as for example the \textit{randn} initialization scheme demonstrates.

\subsection{BP Convergence on non-sign-preserving Tensor Networks}
\label{sec:app:practice:nondef}

We have discussed in Secs.~\ref{sec:backgr:BP:QMBP} and \ref{sec:BPDMRG:alg:nondef} that the BP iteration may also be executed on expectation values $\braket{\psi|H|\psi}$, and that it indeed converges to accurate contraction results if $H$ is a definite operator. If this is not the case, i.e., if $H$ has both positive and negative eigenvalues, BP convergence is not guaranteed, since the TN is not sign-preserving. We demonstrate this by creating different operators $H$ randomly, and executing BP for $\braket{\psi|H|\psi}$. We choose to sample random Pauli strings $P$ with varying weights, such that the impact of non-definite operators with different supports may be examined.

Let $\ket{\psi}$ be a TNS with random site tensors $T_a$, where the underlying lattice is a heavy-hexagonal graph with $2\times 2$ unit cells. This graph has 35 nodes. Pauli strings are defined as $P=\bigotimes_{a\in\sigma}O_a$. The combination $\sigma\subseteq V$ is drawn randomly from the set $V$ of nodes. $O$ is one of the Pauli operators: $O_a\in\{X,Y,Z\}\:\:\forall a\in\sigma$; for sites $b\notin\sigma$, we set $O_b=I$. We sample strings $P$ for all weights that the graph can support, i.e., $0\leq |\sigma| \leq 35$.

\vspace{\baselineskip}

\begin{figure*}
    \begin{center}
        \includegraphics[width=\textwidth]{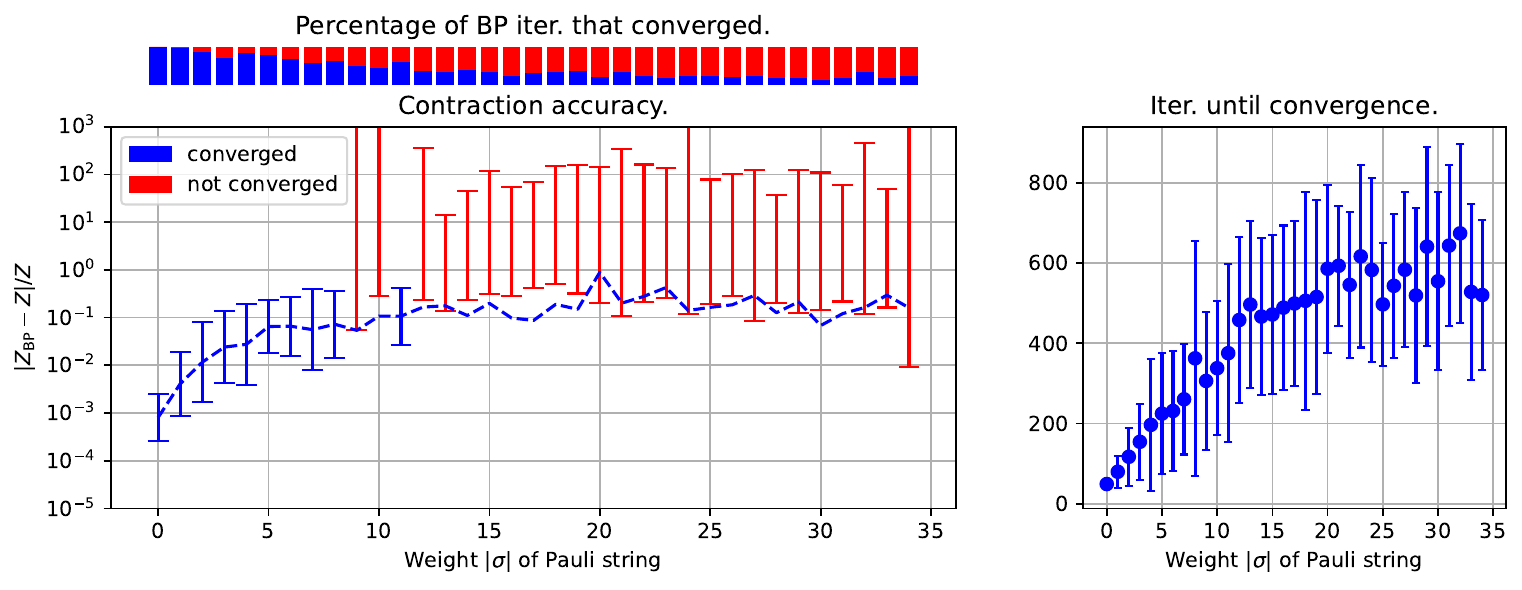}
    \end{center}
    \caption{BP iterations for $\braket{\psi|P|\psi}$, where $\ket{\psi}$ is a TNS on a heavy-hexagonal graph with $2\times 2$ unit cells. The operators $P$ are randomly sampled Pauli strings. For each operator, 50 BP iterations were carried out. For a given weight, the error bar is labeled as ``converged'' if \SI{50}{\percent} or more of the BP iterations converged, and vice-versa. \textbf{(left)} Not only does BP quickly cease to reliably converge for growing Pauli string weight, but the contraction accuracy also decreases significantly. The dashed line shows the mean contraction accuracy when BP does converge. \textbf{(right)} For longer Pauli strings, i.e., more non-definite sites, BP requires more iterations until convergence.}
    \label{app:practice:nondef:fig:non_psd_graphs}
\end{figure*}

Results are displayed in Fig.~\ref{app:practice:nondef:fig:non_psd_graphs}. Clearly, BP ceases to converge reliably. On this graph with 35 nodes, a non-definite operator with $|\sigma|\geq 9$ prevents BP from converging in more than \SI{50}{\percent} of cases.\footnote{Recall that we consider the BP iteration converged if messages do not change upon application of Eq.~\ref{backgr:BP:QMBP:eq:msg_update} (see Sec.~\ref{sec:backgr:BP:TN}).} Contraction accuracy also decreases drastically; for converged BP iterations on $\abs{\sigma}\geq 9$, $\left\langle\log_{10}\abs{Z_\text{BP}-Z}/Z\right\rangle\approx -1$.

Thus, BP must be employed with caution on non-definite graphs. Note, however, that $\left\langle\log_{10}\abs{Z_\text{BP}-Z}/Z\right\rangle\approx -1$ in the cases when BP does converge. If a non-definite graph is encountered in practical applications, BP can be restarted until convergence is achieved, and a crude approximation of the contraction value $Z$ can still be recovered.

\end{document}